\documentclass[10pt]{elsarticle}

\usepackage{blindtext}
\usepackage{fullpage}
\usepackage{graphicx, float}
\usepackage{soul}
\usepackage{amsmath}
\usepackage{amsfonts}
\usepackage{xcolor} 
\usepackage{caption}
\definecolor{darkgreen}{rgb}{0.0, 0.39, 0.0}
\definecolor{darkblue}{rgb}{0.0, 0.0, 0.55}
\usepackage{tikz}
\usepackage[caption=false]{subfig}
\usepackage{caption}
\usepackage{booktabs}
\newsavebox{\mytikzbox}
\sbox{\mytikzbox}{%
    \tikz{\draw[blue, dashed, line width=2pt] (0,0) -- (6mm,0);}%
}
\biboptions{sort&compress}

\usepackage[ruled,vlined]{algorithm2e}
\usepackage{csquotes}

\definecolor{myBlueLink}{RGB}{70,70,200}  
\definecolor{myBlue}{RGB}{18,28,43}  
\definecolor{myOrange}{RGB}{26,82,126} 
\definecolor{IllinoisOrange}{RGB}{232,74,39}  
\definecolor{IllinoisBlue}{RGB}{19,41,75}   
\definecolor{mainText}{RGB}{18,28,43}  

\definecolor{ForestGreen}{rgb}{0.0, 0.2, 0.13}
\definecolor{darkpastelgreen}{rgb}{0.01, 0.75, 0.24}
\definecolor{antiquewhite}{rgb}{0.98, 0.92, 0.84}

\def \pp#1#2{\frac{\partial #1}{\partial #2}}

\newcommand{\ol}[1]{\overline{#1}}
\newcommand{\wt}[1]{\widetilde{#1}}

\newcommand{\bx}{\mathbf{x}}
\newcommand{\bu}{\mathbf{u}}

\newcommand{\bh}{\mathbf{h}}

\newcommand{\DNSsub}{{\text{\textsc{dns}}}}

\def \Da {\mathrm{Da}}

\def \Das {\mathrm{Da}_s}
\def \Ka {\mathrm{Ka}}

\def \Rey {\mathrm{Re}}
\def \Ma {\mathrm{Ma}}
\def \Pr {\mathrm{Pr}}
\def \Sc {\mathrm{Sc}}
\def \Le {\mathrm{Le}}

\newcommand{\keps}{{$k$--$\epsilon$}}

\definecolor{lc1}{HTML}{CE282A}
\definecolor{lc2}{HTML}{255BCB}
\definecolor{lc3}{HTML}{009647} 
\definecolor{lc4}{HTML}{A21D8E}
\definecolor{lc5}{HTML}{F1AB65}
\definecolor{lc6}{HTML}{89B6D4}
\definecolor{lc7}{HTML}{A0A0A0}

\DeclareMathOperator*{\argmin}{argmin}

\def \pp#1#2{\frac{\partial #1}{\partial #2}}
\def \Rey {\mathrm{Re}}
\def \Ma {\mathrm{Ma}}
\def \Pr {\mathrm{Pr}}
\def \Sc {\mathrm{Sc}}
\def \Le {\mathrm{Le}}
\def \Da {\mathrm{Da}}
\journal{arXiv} 

\begin{document}

\begin{frontmatter}
\title{Neural network-augmented eddy viscosity closures for turbulent premixed jet flames}

\author[1]{Priyesh Kakka}
\ead{pkakka@nd.edu}

\author[1]{Jonathan F. MacArt\corref{cor1}}
\ead{jmacart@nd.edu}
\cortext[cor1]{Corresponding author}

\address[1]{Department of Aerospace and Mechanical Engineering, University of Notre Dame, Notre Dame, IN, USA}

\begin{abstract}
Extending gradient-type turbulence closures to turbulent premixed flames is challenging due to the significant influence of combustion heat release.
We incorporate a deep neural network (DNN) into Reynolds-averaged Navier--Stokes (RANS) models for the turbulent viscosity and thermal conductivity as nonlinear functions of the local flow state and thermochemical gradients. Our models are optimized over the RANS partial differential equations (PDEs) using an adjoint-based data assimilation procedure. Because we directly target the RANS solution, as opposed to the unclosed terms, successfully trained models are guaranteed to improve the in-sample accuracy of the DNN-augmented RANS predictions. We demonstrate the learned closures for in- and out-of-sample \emph{a posteriori} RANS predictions of compressible,  premixed, turbulent jet flames with turbulent Damk\"ohler numbers spanning the gradient- and counter-gradient transport regimes. The  DNN-augmented RANS predictions have one to two orders of magnitude lower spatiotemporal mean-squared error than those using a baseline $k$--$\epsilon$ model, even for Damk\"ohler numbers far from those used for training. This demonstrates the accuracy, stability, and generalizability of the PDE-constrained modeling approach for turbulent jet flames over this relatively wide Damk\"ohler number range.
\end{abstract}


\end{frontmatter}

\noindent
\textbf{Novelty and Significance}

\vspace{0.5em}
\noindent
We develop a deep learning turbulence closure method for RANS calculations of turbulent premixed flames. The closure method embeds an untrained neural network into the RANS equations and then optimizes it over the flow solution using an adjoint-based technique. {Novelty:} This is the first application of solver-embedded deep learning to turbulent premixed jet flames. {Significance:} The method is a new, general-purpose closure-modeling framework for turbulent flames. For the present turbulent premixed jet flames, the method significantly reduces the error of \emph{a posteriori} RANS predictions. The trained models generalize this improved accuracy across a wide range of Damköhler numbers, even in combustion regimes that are far out-of-sample from those used to train a particular model, which is not typical of deep learning closures.
\vspace{0.75em}

\noindent
\textbf{Author Contributions}
\begin{itemize}
\item Priyesh Kakka: conceptualization, methodology, software, validation, formal analysis, investigation, data curation, writing - original draft, visualization
\item Jonathan F.\ MacArt: conceptualization, software, resources, writing - review \& editing, supervision, project administration, funding acquisition
\end{itemize}
\vspace{-1em}

\section{Introduction}
\label{sec:Intro}
Navier--Stokes solutions in the turbulent regime are required to design airfoils, nozzles, heat sinks, combustors, and many other engineering devices. The computational effort required to predict turbulent flows with high-fidelity direct numerical simulation (DNS) is prohibitively expensive due to its cost scaling with $\Rey^3$~\cite{pope2000turbulent}. Therefore, turbulent flows are usually solved with partially or fully modeled turbulence scales. Two of the most common modeling techniques are the large-eddy simulation (LES) and Reynolds-averaged Navier--Stokes (RANS) equations. In LES, length scales in the dissipation range are truncated and modeled, while the remaining larger scales are resolved. In RANS, all turbulence scales are modeled, yielding spatially and/or temporally averaged solutions. For both methods, the reduced-order governing partial differential equations (PDEs) contain unclosed terms that represent the ``missing'' physics. Historically, closure models for these have been developed for specific configurations such as nonreacting free shear flows~\cite{yoder2015modeling} and wall-bounded flows~\cite{piomelli2008wall, bose2018wall}, among many others, which limits their predictive accuracy for reacting flows.

Most closure models for conventional RANS and LES utilize the Boussinesq hypothesis for the Reynolds stresses and the gradient-diffusion hypothesis for the scalar fluxes. Both make restrictive assumptions about the nature of turbulence that limit their utility for reacting flows~\cite{frank1999measurements,veynante1997gradient,macart2018effects}. The Boussinesq hypothesis for RANS (LES) requires that the (subfilter) turbulent dissipation locally balances the (subfilter) turbulence production and further assumes the alignment of the Reynolds stress (subfilter stress) anisotropy with the mean (filtered) strain-rate tensor. The gradient-diffusion hypothesis assumes that the (subfilter) scalar flux vectors are aligned with the respective mean (filtered) scalar gradients~\cite{pope2000turbulent}.

RANS modeling for turbulent premixed combustion is complicated by the presence of regime-dependent flame--turbulence interactions. Chemical heat release causes local volumetric expansion, which occurs roughly on the scales of the flame thickness and leads to flame-induced pressure dilatation~\cite{zhang1995premixed,o2017cross}. The Boussinesq and gradient-diffusion hypotheses fail if the relative influence of flame-induced pressure dilatation is greater than that of turbulence-induced flow straining \cite{macart2018effects}. The dominance of these two effects can be parameterized on the relative time scales of the flame and the small-scale turbulence---the Karlovitz number~\cite{peters2001turbulent},
\begin{equation}
    \label{eq:Ka_no}
    \Ka \equiv \frac{t_{F}}{t_{\eta}}=\frac{\delta_{F}}{ s_{L}}\left(\frac{\epsilon}{ \nu}\right)^{1 / 2} ,
\end{equation}
where $t_{F}$ is the laminar flame time scale, $t_{\eta}$ is the Kolmogorov time scale, $s_{L}$ is the laminar flame speed, $\epsilon$ is the turbulent kinetic energy (TKE) dissipation rate, $\nu$ is the kinematic viscosity, and $\delta_{F}$ is the thickness of the corresponding 1D laminar flame~\cite{poinsot2005theoretical}
\begin{equation}
\label{eq:flame thickness}
    \delta_F = \frac{T_2 - T_1}{\max \left( \left| \frac{\partial T}{\partial x} \right| \right)},
\end{equation}
where $T_2$ and $T_1$ are the temperatures in the products and reactants, respectively, and $\max \left( \left| \frac{\partial T}{\partial x} \right| \right)$ is the maximum temperature gradient within the flame.

At low Karlovitz numbers and high Damk\"ohler numbers (i.e., the thin-flames regime~\cite{peters2001turbulent}), flame-induced pressure dilatation dominates turbulence-induced strain~\cite{zhang1995premixed,veynante1997gradient,macart2018effects}. Thin flames are common in high-pressure regions of combustors, where flame thickness decreases with increasing pressure; hence, accurate modeling of thin flames is crucial.
DNS analyses of turbulent, premixed, planar jet flames have shown pressure-dilatation to induce high levels of Reynolds stress anisotropy in the thin-flames regime \cite{macart2018effects,macart2019evolution,macart2021damkohler}. These studies found a concomitant reversal of the balance of turbulence production and dissipation at high Damk\"ohler numbers, with the velocity-pressure gradient correlation term becoming the largest source of TKE, and the turbulence production term becoming a stronger sink of TKE than the viscous dissipation. These findings illustrate the failure of the Boussinesq and gradient-diffusion hypotheses in these regimes, as it presumes the TKE production  balances dissipation~\cite{pope2000turbulent}. Moreover, the strong velocity-pressure gradient correlation term in high-Damk\"ohler number turbulent premixed flames leads to misalignment between the mean strain rate and the Reynolds stresses, which voids the algebraic form of the Boussinesq model.

A large neural network can act as a universal approximator of any function~\cite{hornik1989multilayer}. Neural networks thus allow nonlinear functional mappings to be obtained from known data and/or a set of constraints. Different approaches have incorporated neural networks to predict turbulent flows and other physically challenging problems. One approach involves surrogate models trained using high-fidelity simulation data. Such surrogate models can be trained to predict flow-field quantities such as unresolved and/or unclosed terms in turbulent flows. Ling \emph{et al}.~\cite{ling2016reynolds} formulated a neural network architecture to preserve the Galilean invariance of anisotropy tensors which were tested \emph{a priori} as RANS models. Similar attempts to develop LES models have used~\textit{a priori} training of neural networks~\cite{maulik2017neural} or superresolution models~\cite{liu2020deep}, including several superresolution methods for reacting flows \cite{bode2021using,Bode2023,nista2023investigation,nista2024influence}. Comprehensive reviews of data-driven machine learning for turbulence modeling in nonreacting flows have been given by Brunton \emph{et al.}~\cite{brunton2020machine} and Duraisamy \emph{et al.}~\cite{duraisamy2019turbulence}. Another approach is to train neural networks using the governing equations, after which the predictions by the surrogate models are made without the help of any equations, which are usually called physics-informed neural networks (PINNs)~\cite{raissi2019physics}. These latter methods, while attractive, have inherent issues with their training, especially for complex problems such as the unsteady Navier--Stokes equations~\cite{wang2021understanding}.

Deep learning (DL) modeling of reduced-order combustion kinetics has attracted high levels of interest. Perry \emph{et al.}~\cite{perry2022co} obtained reduced-order manifolds for chemical kinetics using a neural network structure that simultaneously encodes the chemistry manifold variables, their nonlinear mapping, and the LES subfilter closures. Related work for kinetics modeling has been carried out by Ji \emph{et al.}~\cite{ji2021stiff}, Owoyele and Pal~\cite{owoyele2022chemnode}, and others. Conversely, the application of neural networks for the turbulent transport terms is relatively unexplored. Yellapantula \emph{et al.}~\cite{yellapantula2021deep} showed \emph{a priori} tests of DL models to predict the subfilter contribution to the progress variable dissipation rate, and Lapeyre \emph{et al.}~\cite{lapeyre2019training} employed convolutional neural networks to predict flame surface density. These efforts were solely data-driven for training the neural networks, i.e., without incorporating explicit PDE constraints.  Integrating DL models for $\textit{a posteriori}$ simulations of turbulent combustion, and using PDE constraints while training neural networks, remain a challenge.

To incorporate neural networks into numerical simulations of dynamical systems, trusted data can be incorporated into the  physics solver using a deep learning PDE augmentation method (DPM)~\cite{sirignano2020dpm}. This method employs a memory-efficient, adjoint-based optimization algorithm and neural networks to enforce the model-consistent constraint of the PDEs~\cite{duraisamy2021perspectives}. The deep learning PDE augmentation method has also demonstrated potential for accuracy in out-of-sample cases. For instance, MacArt \emph{et al.}~\cite{macart2021embedded} trained an LES model on a turbulent planar single jet that was accurate for \emph{a posteriori} predictions of geometrically out-of-sample interacting jet pairs. Similarly, DPM has been used for LES predictions of bluff-body wakes for out-of-sample aspect ratios, blockage shapes, and Reynolds numbers~\cite{sirignano2023deep}, steady RANS predictions of turbulent channel flows at out-of-sample Reynolds numbers~\cite{sirignano2023pde}, and compressible hypersonic flows in the transition-continuum regime at out-of-sample Mach numbers~\cite{nair2023deep}. However, these previous studies all focused on nonreacting flows. The objective of the present work is to develop a generalized turbulence model using DPM for compressible, turbulent, reacting flows. We approach this by enhancing existing RANS models to be applicable to turbulent reacting flows.

As a first step, we focus on RANS predictions of turbulent premixed jet flames due to the inherently lower PDE-constrained optimization (training) cost of RANS compared to LES. The optimization method is more general than the PDEs being solved and could be applied equally to LES, as has been done previously for nonreacting flows, though at concomitantly higher cost. To generate target data for model optimization, we simulate temporally evolving turbulent premixed jet flames spanning a range of Damk\"ohler numbers using DNS. Baseline compressible RANS predictions of these flames use a standard $k$--$\epsilon$ closure.

To provide a well-constrained foundation for DL optimization, we design a neural network closure that augments the baseline \keps-modeled eddy viscosity and thermal conductivity. This approach, while functionally more limited than direct closure of the Reynolds stress and scalar fluxes using neural networks, is mathematically simpler to constrain for physical realizability (e.g., invariance properties) and is more straightforward to train, as the baseline model provides a reasonable starting point. Augmenting an existing model is also more interpretable, as the baseline model tends to be better understood than a neural network alone, and is also potentially more generalizable to out-of-sample conditions, as the additional physical constraints can help to avoid overfitting. The choice of baseline model is not unique; any turbulence model (e.g., a wall-adapting model for wall-bounded flows \cite{hickling2024large,liu2024adjoint}) could be augmented in the same way using embedded optimization (DPM). 
We use single-step global kinetics in order to focus on the turbulent transport terms, though in principle the DPM approach could be applied to any kinetic description. We assess the accuracy of DL-augmented \emph{a posteriori} predictions across a wide range of in- and out-of-sample Damk\"ohler numbers.

The article is structured as follows. Section~\ref{sec:Problem formulation} introduces the governing equations for DNS and RANS and details the turbulent premixed jet flame configuration. Section~\ref{sec:DPM} presents the DPM approach and training methodology. Model capabilities are evaluated in Section~\ref{sec:Results}. We  discuss the computational cost, advantages, and challenges of the DPM approach in Section~\ref{sec:discussion}. Finally, conclusions are provided in Section~\ref{sec:conclusion}.

\section{Governing equations and numerical methods}\label{sec:Problem formulation}

This section presents our mathematical framework and numerical methods for DNS and RANS of turbulent premixed jet flames. Section~\ref{subsec:gov_eqn}  introduces the dimensionless governing equations for compressible reacting flows, Section~\ref{subsec:DNS_formulation} outlines the computational setup and parameters employed in DNS, Section~\ref{sec:avg_var} presents DNS validation results, Section~\ref{subsec:RANS_formulation} provides the RANS governing equations, and Section~\ref{sec:eddyvisc} introduces the RANS neural network closures.

\subsection{Governing equations and constitutive relations}\label{subsec:gov_eqn}
We model turbulent premixed planar jet flames in which a central slot jet issues a fuel--air premixture into quiescent surroundings of hot products.  To focus on the turbulent transport terms, we describe the chemical kinetics using a global, irreversible reaction with $N=2$ species calibrated to approximate stoichiometric H$_2$--air combustion, diluted by 30\,\% volume with N$_2$ \cite{bane2010development}. The second-order global reaction is
\begin{equation}
\label{eq:reaction}
  2R \rightarrow 2P,
\end{equation}
where \( R \) and \( P \) represent the reactant and product species. The use of a global reaction enables the heat release effects on turbulence to be quantified by a single dimensionless parameter, the scaling Damk\"ohler number, \( \Da_s \).
For simplicity of the  subsequent governing equations, we define both species to possess the elemental properties of argon, which results in constant and equal species molecular weights \( \hat{W}_k \), species specific heats at constant pressure \( \hat{C}_{p,k} \), and ratios of specific heats \( \gamma_k \). Further, because the species have equal molecular weights, the species mole fractions \( X_k \) are equal to the respective species mass fractions \( Y_k \). The progress variable is equivalent to the product species mass fraction, \( Y_P \).

The compressible, reacting Navier--Stokes equations are nondimensionalized following Wang \emph{et al.}~\cite{wang2021flow}. Dimensional reference values comprise a characteristic length \(\hat{L}_{0}\), 
velocity \(\hat{u}_{0}\), density \(\hat{\rho}_{0}\), pressure \(\hat{p}_{0}\), temperature \(\hat{T}_{0}\), and mean molecular weight \(\hat{W}_0\). All reference flow properties are the initial bulk properties in the reactants jet. This leads to dimensionless variables
\begin{align}
  t = \frac{\hat{t} \hat{u}_{0}}{\hat{L}_0}, \quad
  \mathbf{x} = \frac{\hat{\mathbf{x}}}{\hat{L}_0}, \quad 
  \rho = \frac{\hat{\rho}}{\hat{\rho}_{0}}, \quad
  \mathbf{u} &= \frac{\hat{\mathbf{u}}}{\hat{u}_{0}}, \quad
  {p} = \frac{{\hat{p}}}{\gamma \hat{p}_{0}}, \quad
  T = \frac{\hat{T}}{(\gamma-1)\hat{T}_{0}}, \nonumber \\
  e = \frac{\hat{e}}{\hat{u}_{0}^{2}}, \quad
  W_k &= \frac{\hat{W}_{k}}{\hat{W}_0}, \quad
  h_{k} = \frac{\hat{h}_{k}\hat{W}_{k}}{\hat{R}_{u}\hat{T}_{0}},
\label{eq:non-dim}
\end{align}
where $t\geq 0$ and $\bx\in\mathbb{R}^3$ are the dimensionless time and space coordinates, $\bu\in\mathbb{R}^3$ is the dimensionless velocity vector, $e$ is the dimensionless mixture internal energy,  ${h}_k$ are the dimensionless species enthalpies incorporating sensible and chemical energies,  $\gamma$ is the uniform mixture ratio of specific heats,  and \(\hat{R}_u = 8.314\) $\mathrm{J/(mol}\cdot\mathrm{K})$ is the universal gas constant. 
The resulting dimensionless governing equations are
\begin{align}
  \frac{\partial \rho}{\partial t} &+ \frac{\partial \rho u_j}{\partial x_j} = 0, \label{eq:NS_cont} \\
  \frac{\partial \rho u_i}{\partial t} &+ \frac{\partial \rho u_i u_j}{\partial x_j} + \frac{1}{\Ma^2}\frac{\partial {p}}{\partial x_i} - \frac{1}{\Rey}\frac{\partial \tau_{ij}}{\partial x_j} = 0, \label{eq:NS_mom} \\
  \frac{\partial \rho E}{\partial t} &+ \frac{\partial \rho u_j E}{\partial x_j} + \frac{1}{\Ma^2}\frac{\partial p u_j}{\partial x_j} - \frac{1}{\Rey}\frac{\partial u_i \tau_{ij}}{\partial x_j} - 
  \frac{1}{\Ma^2 \Rey \Pr}\frac{\partial}{\partial x_j}\left({\mu}\frac{\partial T}{\partial x_j}\right) \notag \\
  &- \frac{ 1}{\gamma \mathrm{Ma}^2 \operatorname{RePrLe}} \frac{\partial}{\partial x_j}\left(\rho \sum_{k=1}^N \frac{Y_k}{W_k} h_k {V}_{k, j}\right)=0, \label{eq:NS_energy}\\
  \frac{\partial \rho Y_{P}}{\partial t} &+ \frac{\partial}{\partial x_{j}} \left(\rho Y_{P} u_{j}\right) - \frac{1}{\Rey \Pr \Le} \frac{\partial}{\partial x_{j}} \left(\rho Y_{P} V_{P,j}\right) - \Da_{s} W_{P} \dot{\omega}_{P} = 0, \label{eq:NS_species}
\end{align}
where \({E} = {e} + u_iu_i/2\) is the total energy, and $\Rey$, $\Ma$, $\Pr$, and $\Le$ are the scaling Reynolds, Mach, Prandtl, and Lewis numbers, values of which are provided in Section~\ref{subsec:DNS_formulation} for the present turbulent premixed jet flames.
The viscous stress tensor and the mass diffusion velocity of species $k$ are modeled by
\begin{align}
  \tau_{ij} &= {\mu} \left(\frac{\partial u_i}{\partial x_j} + \frac{\partial u_j}{\partial x_i} - \frac{2}{3}\frac{\partial u_k}{\partial x_k}\delta_{ij}\right),~\label{eq:stress_tensor} \\
 V_{k,j} &= \mu\left(\frac{1}{Y_{k}} \frac{\partial Y_{k}}{\partial x_{j}} \right),~\label{eq:diff_vel} 
\end{align}
where $\delta_{ij}$ is the Kronecker delta, and summation over the species index in \eqref{eq:diff_vel} is not implied. We model the dimensionless base viscosity using a power-law dependence: \({\mu} = \hat{\mu}/{\mu_{0}} = \left[(\gamma - 1) {T}\right]^{0.7}\). 

The progress variable net production rate is
\begin{align}
  \dot{\omega}_{P} &=  \nu_{P}  \left(\frac{\rho (1 - Y_{P})}{W_{R}}\right)^{\nu_{P}} T^{\beta} \exp{\left(-\frac{T_{a}}{(\gamma - 1) T }\right)},\label{eq:Omega}
\end{align}
where \(\nu_{P} = 2\) is the molar stoichiometric coefficient from \eqref{eq:reaction}, and \(\beta = 0\) is the temperature exponent.
The dimensionless activation energy is \(T_{a} = {\hat{E}_{a}}/{(\hat{{R}}_u \hat{T}_{0})} = 46.72\) with \(\hat{E}_{a} = 27{,}856\)~kcal/mol to match experimentally observed ignition delay times \cite{bane2010development}.
The scaling Damk\"ohler number is
\begin{equation}
\label{eq:Da}
    \Da_{s} = \hat{W}_{P}\frac{\hat{A}\hat{L}_0 (\hat{T}_{0}(\gamma-1))^{\beta}}{\hat{\rho}_{0}\hat{u}_{0}}\left(\frac{\hat{\rho}_{0}}{\hat{W}_{P}}\right)^{\nu_{P}},
\end{equation}
where $\hat A$ is an Arrhenius pre-exponential factor. The utility of the dimensionless formulation lies in the ability to prescribe $\Da_s$ directly rather than needing to evaluate it from \eqref{eq:Da} for prescribed dimensional conditions. Further details on the range of $\Da_s$ simulated are provided in Section~\ref{subsec:DNS_formulation}.

Finally, the dimensionless equation of state is
\begin{equation}
\label{eq:eqn_state}
  p = \frac{(\gamma-1)}{\gamma W} \rho T,
\end{equation}
in which the dimensionless temperature is evaluated from the caloric equation of state
\begin{equation}
\label{eq:int_energy}
    e(T,Y_k) = \left( \frac{1}{\gamma \Ma^2}\left(\sum_{k=1}^{N}\frac{Y_{k}}{W_{k}}\left(h_{k}(T) -\left(\gamma - 1\right)T  \right)\right)\right),
\end{equation}
where the species enthalpies \( h_k(T) \) are evaluated using dimensionless polynomials calibrated for the global reaction \cite{bane2010development},
\begin{align}
  \label{eq:enthap_reac}
  \begin{split}
    {h}_R(T) &=  \frac{W_{R}}{\hat{T}_0}\left( 2.5 (\gamma -1 )\hat{T}_0{T} - 745.375 \right), \\
    {h}_P(T) &= \frac{W_{P}}{\hat{T}_0} \left( 2.5 (\gamma -1 )\hat{T}_0{T} - 5972 \right).
  \end{split}
\end{align}
From \eqref{eq:int_energy} and \eqref{eq:enthap_reac}, the dimensionless temperature is
\begin{equation}
T = \frac{e \gamma \Ma^2 \hat{T}_0 + 745.375 (1 - Y_P) + 5972 Y_P}{1.5 (\gamma - 1) \hat{T}_0}.
\label{eq:dimensionless_caloric_eos}
\end{equation}
This methodology yields adiabatic flame temperatures comparable to those obtained from the detailed mechanism of Li \emph{et al.}~\cite{li2004updated}. Using this configuration, Bane \emph{et al.}~\cite{bane2010development} found the global-mechanism flame speed to be within 6\,\% of the detailed-mechanism flame speed and the global-mechanism adiabatic flame temperature to be within 2\,\% of that of the detailed mechanism.

The system of equations is solved using a high-performance, fully graphics processing unit (GPU)-accelerated, Python-native flow solver, \textit{PyFlowCL} \cite{liu2024adjoint, hickling2024large}. The solver uses fourth-order central-difference schemes  on structured curvilinear meshes, fourth-order explicit Runge--Kutta time advancement, and eighth-order compact filters~\cite{lele1992compact} to alleviate spurious oscillations. \emph{PyFlowCL} leverages the \emph{PyTorch}~\cite{paszke2019pytorch} high-performance linear algebra library for neural network implementation and algorithmic differentiation, which we used to construct the adjoint equations needed for optimization.

\subsection{Direct numerical simulations}
\label{subsec:DNS_formulation}
DNS datasets are generated for temporally evolving turbulent premixed planar jet flames at  $\Rey=6{,}000$,  $\mathrm{Ma}=0.08$, \(\Pr = 0.7\), and \(\Le= 1.14\). The simulation domain has dimensionless extents $(L_x, L_y, L_z)=(16H, 30H, 12H)$ in the streamwise direction $x$, cross-stream direction $y$, and spanwise direction $z$, respectively, where $H=1$ is the initial slot-jet height (the characteristic length scale). The streamwise and spanwise directions are periodic and spatially homogeneous, with absorbing layers applied along the cross-stream boundaries \cite{liu2024adjoint}.

We discretize the DNS domain using $(N_x,N_y,N_z)=(1024, 800, 768)$ mesh points spaced uniformly in the streamwise and spanwise directions and stretched in the cross-stream direction according to the mapping
\begin{equation}
y_j = \left( 1 + \frac{\sinh\left( \delta_{y} \left( {\eta_j} - {1}/{2} \right) \right)}{\sinh({\delta_{y}}/{2})} \right) \frac{L_{y}}{2}, \quad j\in[1,N_y],
\end{equation}
where $\eta_j\in[0,1]$ is the uniform computational mesh coordinate in the cross-stream direction, and  $\delta_y = 6$ is a mesh stretching factor. This mesh stretching ensures ${\tilde{\eta}}/{\Delta}_\DNSsub\geq 0.5$, where $\tilde{\eta}$ is the Kolmogorov length scale and $\Delta_\DNSsub$ is the local DNS mesh resolution (geometric mean), and $\delta_F/\Delta_\DNSsub \geq 5$, both of which are generally considered sufficient for turbulent premixed combustion DNS.
The time step for DNS is $\Delta t=3.75 \times 10^{-4}$, which corresponds to a typical Courant--Friedrichs--Lewy (CFL) number of 0.39.

The direct numerical simulations are initialized at \( t = 0 \) with a nonuniform jet core flow:
\begin{equation}
u_j = \frac{1}{2} \left[\tanh\!\left(\frac{y_j/H + 1/2}{\delta_0}\right) - \tanh\!\left(\frac{y_j/H - 1/2}{\delta_0}\right)\right], \quad j \in [1, N_y],
\end{equation}
where the nominal shear layer thickness is \( \delta_0 = 0.05 \). {To initialize turbulence, synthetic fluctuations are superimposed following the method of Passot and Pouquet~\cite{passot1987numerical} while maintaining a turbulence intensity of 4\,\% with respect to the reference velocity. The prescribed fluctuations are obtained from a synthetic energy spectrum
\begin{equation}
\label{eq:tke_spec}
E(k) = C k^4 \exp\left(-\frac{k^2}{k_0^2}\right),
\end{equation}
where the characteristic wavenumber \( k_0 = 3 \) corresponds to a dimensionless integral length scale \( l = 1/3 \). The normalization constant $C = 4.95 \times 10^{-6}$ is obtained by equating the integral of \eqref{eq:tke_spec} to the target TKE. The velocity field is initialized in the frequency domain using \eqref{eq:tke_spec}, after which an inverse Fourier transform is applied to obtain an initial velocity field consistent with the imposed spectrum.

High-temperature products are positioned at \(y=\pm2H\) following the approach of Hawkes \emph{et al.}~\cite{hawkes2012petascale} using the temperature profile
\begin{equation}
\label{eq:temp_profile}
T_{j} = T_P + \frac{T_P - T_0}{2} \left[ \tanh\!\left( \frac{2(y_{j} - 2H)}{\delta_0} \right) - \tanh\!\left( \frac{2(y_{j} + 2H)}{\delta_0} \right) \right], \quad j\in[1,N_y],
\end{equation}
{where $T_P = {\hat{T}_P}/{(\gamma-1)\hat{T}_{0}}$, \(\hat{T}_P = 2396.77\,\text{K}\) is the dimensional adiabatic flame temperature, and \(\hat{T}_{0} = 300\,\text{K}\)  is the dimensional reference temperature.} This configuration ensures that a self-similar jet core develops prior to its ignition (for $\Da_s>0$) upon reaching the hot products \cite{hawkes2012petascale}}. {The species are initialized using an analogous formula to \eqref{eq:temp_profile},
\begin{equation}
\label{eq:Y_profile}
    {Y_P}_{j} = \frac{1}{2} \left[ \tanh\!\left( \frac{2 (y_{j} - 2H)}{\delta_0} \right) - \tanh\!\left( \frac{2 (y_{j} + 2H)}{\delta_0} \right) \right], \quad j\in[1,N_y].
\end{equation}}
The density is initialized using \eqref{eq:eqn_state} with uniform initial pressure $p = 1/\gamma$.

\begin{figure}
    \centering
    \includegraphics[width=0.6\linewidth]{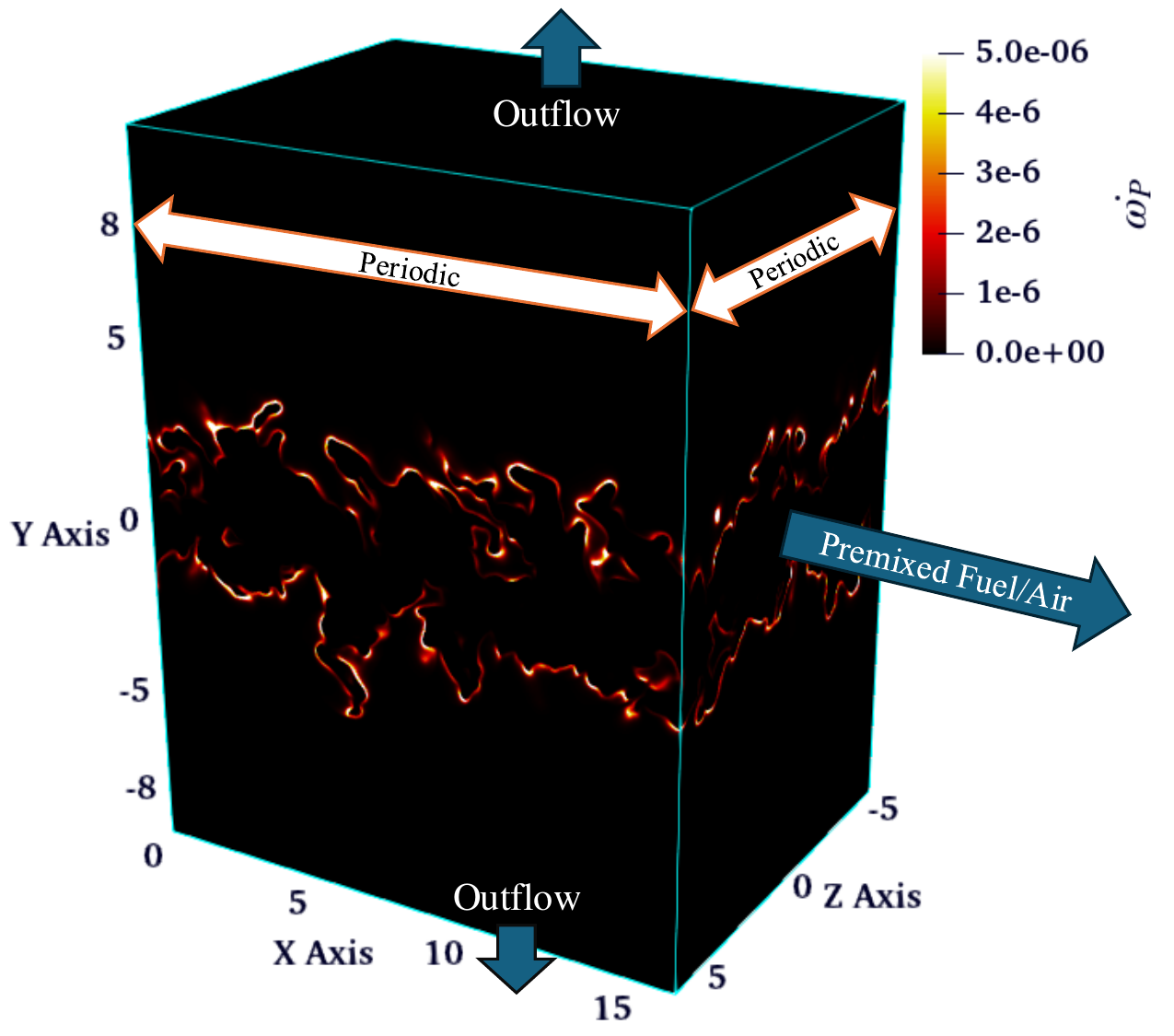}
    \caption{Instantaneous DNS reaction source term for $\Da_{s} =  35{,}000$ at $t=32$. Computational boundary conditions and the flow direction of the jet core are indicated by arrows.}
    \label{fig:flame_domain}
\end{figure}

Six turbulent premixed jet flames are simulated for scaling Damk\"{o}hler numbers $\Da_s\in\{0,6{,}000,13{,}000,$ $20{,}000,27{,}000,35{,}000\}$. 
Turbulent Damk\"{o}hler numbers are evaluated at $Y_{P} = 0.5$ as $\Da_{t} = {s_T \ell}/{\delta_F u^{\prime}}$,
where $s_T$ is the mean DNS-evaluated turbulent flame speed, and $u'$ and $\ell$ are the DNS-evaluated root-mean-square velocity fluctuations and integral length scale.  Table~\ref{tab:conversion} presents the conversion of prescribed scaling Damk\"ohler numbers to DNS-evaluated turbulent Damk\"{o}hler numbers at dimensionless time $t = 32$, which occurs near the end of the DNS calculations when the flame is highly wrinkled (see Figure~\ref{fig:flame_domain}). For \emph{a posteriori} RANS calculations and DPM model optimization, initial conditions are obtained from DNS at $t=15$, at which time the turbulent jet and the flame begin to interact.  DNS data is collected until $t = 33.9$ time units. One DNS dataset is used per scaling Damk\"ohler number.

\begin{table}
\centering
\caption{Prescribed scaling Damk\"ohler numbers and DNS-evaluated turbulent Damk\"{o}hler numbers at $t = 32$.}
\begin{tabular}{lcccccc}
\toprule
$\Da_{s}$ & 0.0 & 6{,}000 & 13{,}000 & 20{,}000 & 27{,}000 & 35{,}000 \\
$\Da_{t,{{Y}_{0.5}}}$ & 0.00 & 0.05 & 0.12 & 0.25 & 0.36 & 0.48 \\
\bottomrule
\end{tabular}
\label{tab:conversion}
\end{table}

\begin{figure}
    \begin{minipage}{0.32\textwidth}
        \includegraphics[width=\linewidth]{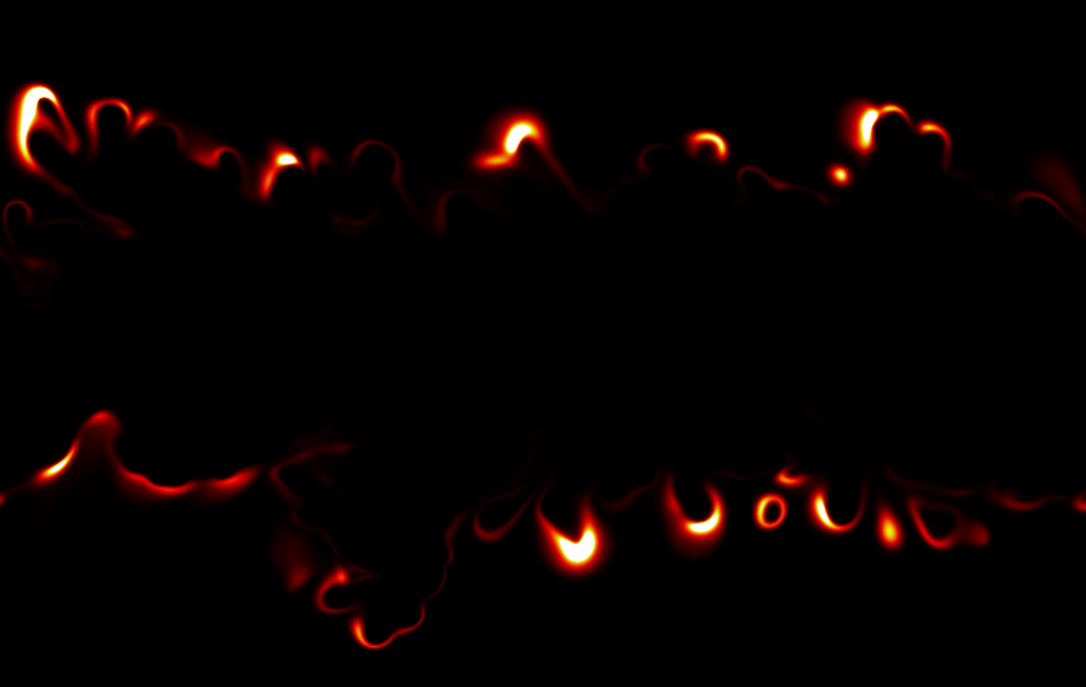}
        \caption*{$\text{Da}_s = 6{,}000$}
    \end{minipage}
    \hfill
    \begin{minipage}{0.32\textwidth}
        \includegraphics[width=\linewidth]{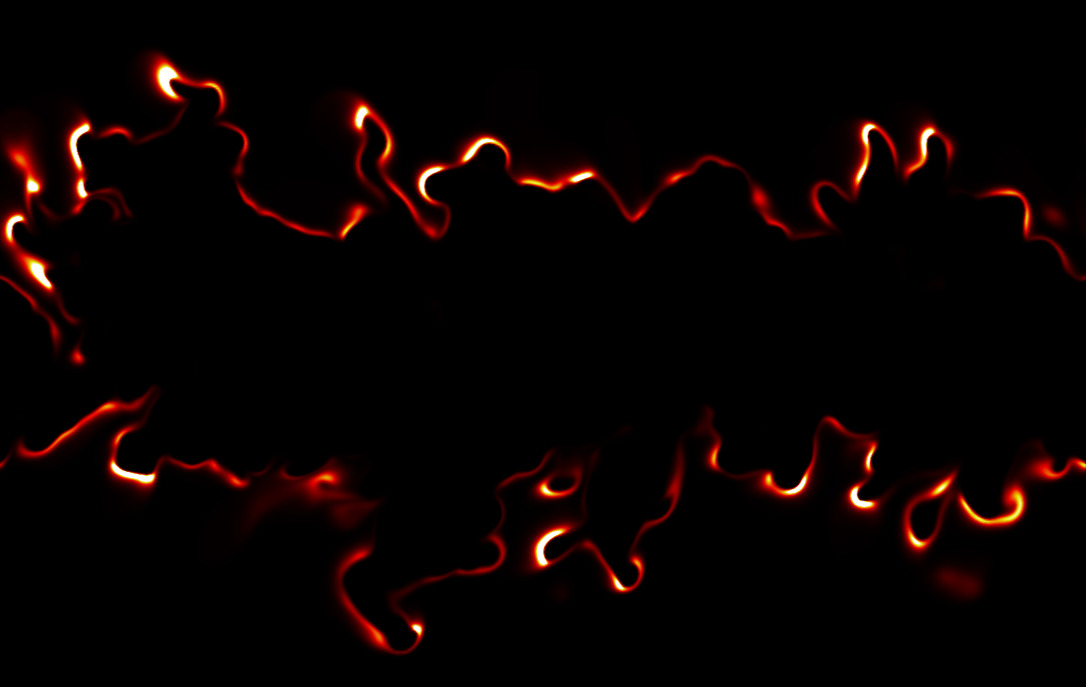}
        \caption*{$\text{Da}_s = 20{,}000$}
    \end{minipage}
    \hfill
    \begin{minipage}{0.32\textwidth}
        \includegraphics[width=\linewidth]{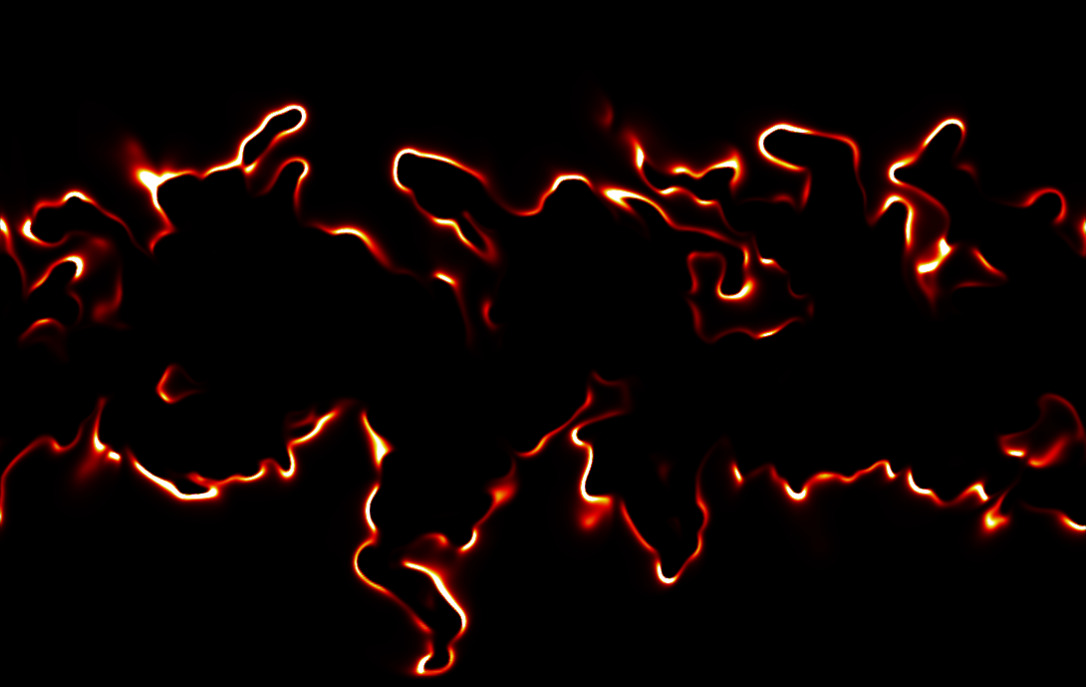}
        \caption*{$\text{Da}_s = 35{,}000$}
    \end{minipage}
    \caption{Instantaneous DNS reaction source
terms for several scaling Damk\"ohler numbers $t = 32$. For interpretation of the color scale, the reader is referred to Figure~\ref{fig:flame_domain}.}
    \label{fig:diff_DA}
\end{figure}

Instantaneous snapshots of the turbulent jet flame DNS for various scaling Damk\"{o}hler numbers are shown in Figure~\ref{fig:diff_DA}. As $\Da_s$ increases, the flame wrinkling intensifies, resulting in an increase in turbulent flame speed. This in turn leads to different interactions between the flame and turbulence.

\subsection{DNS-evaluated flow statistics}
\label{sec:avg_var}
The instantaneous DNS flow variables are decomposed into mean and fluctuating quantities as $\phi=\bar\phi + \phi'$. For the present temporally evolving jet flames,  the Reynolds-averaged quantities $\bar\phi$ are obtained by integrating in the statistically homogeneous $x$ and $z$ directions, yielding spatiotemporal statistics of \(y\) and \(t\):
\begin{equation}
  \ol{\phi}(y_j, t^n) = \frac{1}{N_x N_z} \sum_{i=1}^{N_x} \sum_{k=1}^{N_z}  \phi(x_i, y_j, z_k, t^n).
  \label{eq:RANS_average}
\end{equation}
Favre-averaged quantities are obtained as $\widetilde{\phi} = {\overline{\rho\phi}}/{\bar{\rho}}$
with corresponding fluctuations $\phi^{\prime \prime}=\phi-\widetilde{\phi}$. 

\begin{figure}
\centering
   \begin{minipage}{1\textwidth}
        \centering        \includegraphics[width=0.5\linewidth]{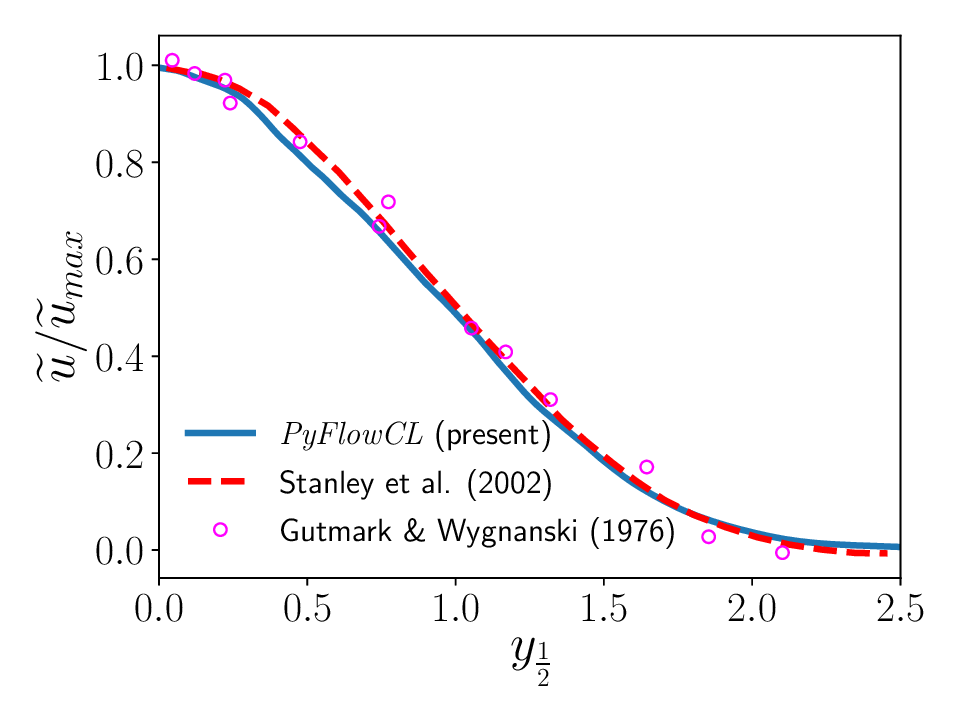}
        \caption{Self-similar scaling of DNS and experimental mean streamwise jet velocity for $\Da_{s}=0$.}
        \label{fig:u_compare}
    \end{minipage}
\end{figure}

The \emph{PyFlowCL} DNS calculations are validated for $\Da_{s} = 0$ by comparing with the DNS of Stanley \emph{et al.}~\cite{stanley2002study} and the experiments of Gutmark \emph{et al.}~\cite{gutmark1976planar}. Figure~\ref{fig:u_compare} compares the mean streamwise velocity $\widetilde{u}$ in self-similar coordinates, where $y_{{1}/{2}}$ is the jet half-width and $\wt{u}_\mathrm{max}$ is the mean centerline velocity, of the three data sets. The \emph{PyFlowCL} mean flow demonstrates good agreement with the published DNS and experiment. Other flow statistics, not reported here, demonstrate similarly close agreement.

\begin{figure}
    \centering
    \hspace{1.1cm} {\small $\Da_s=0$} \hspace{6.85cm} {\small$\Da_s=35{,}000$} \\
    \includegraphics[width=0.49\linewidth]{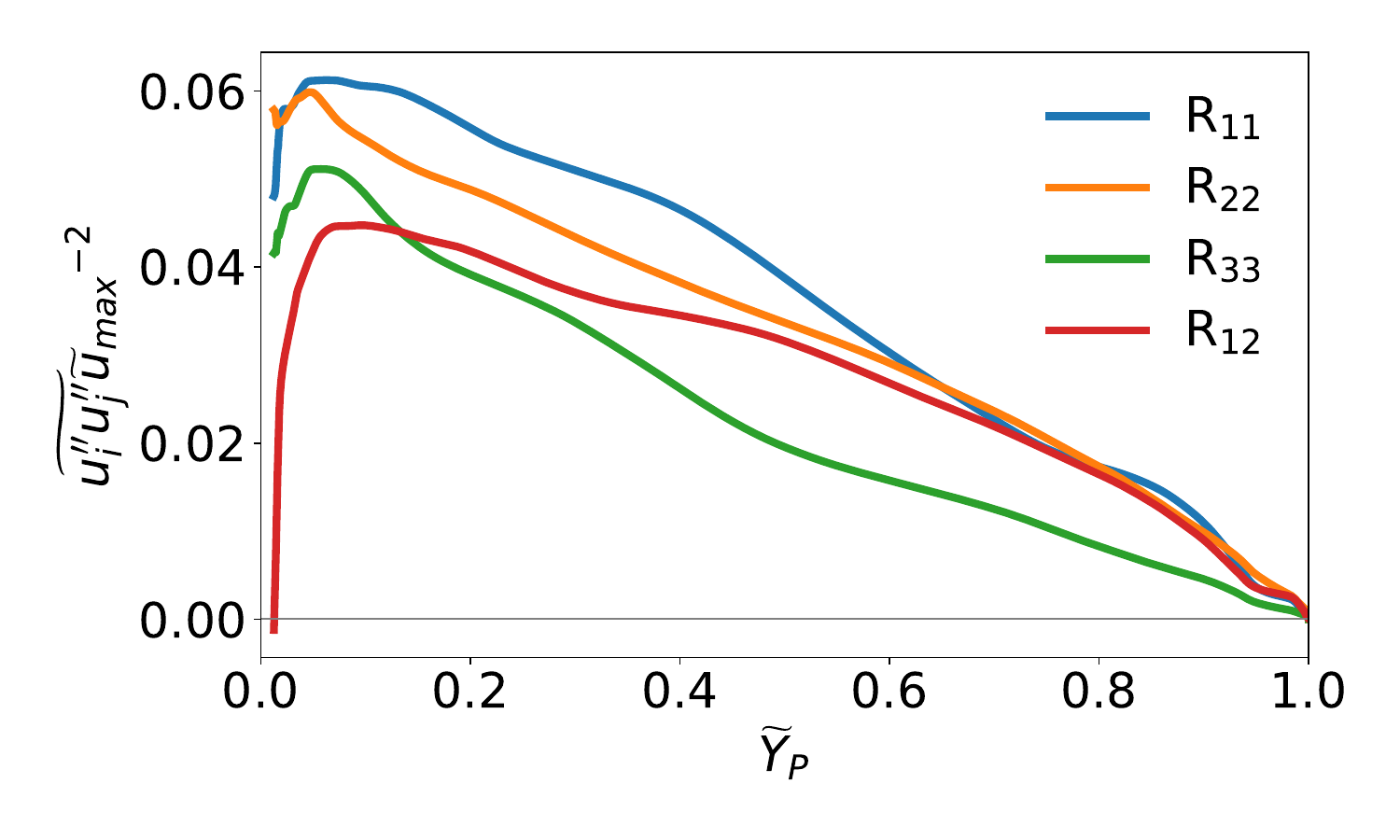}
    \hfill 
    \includegraphics[width=0.49\linewidth]{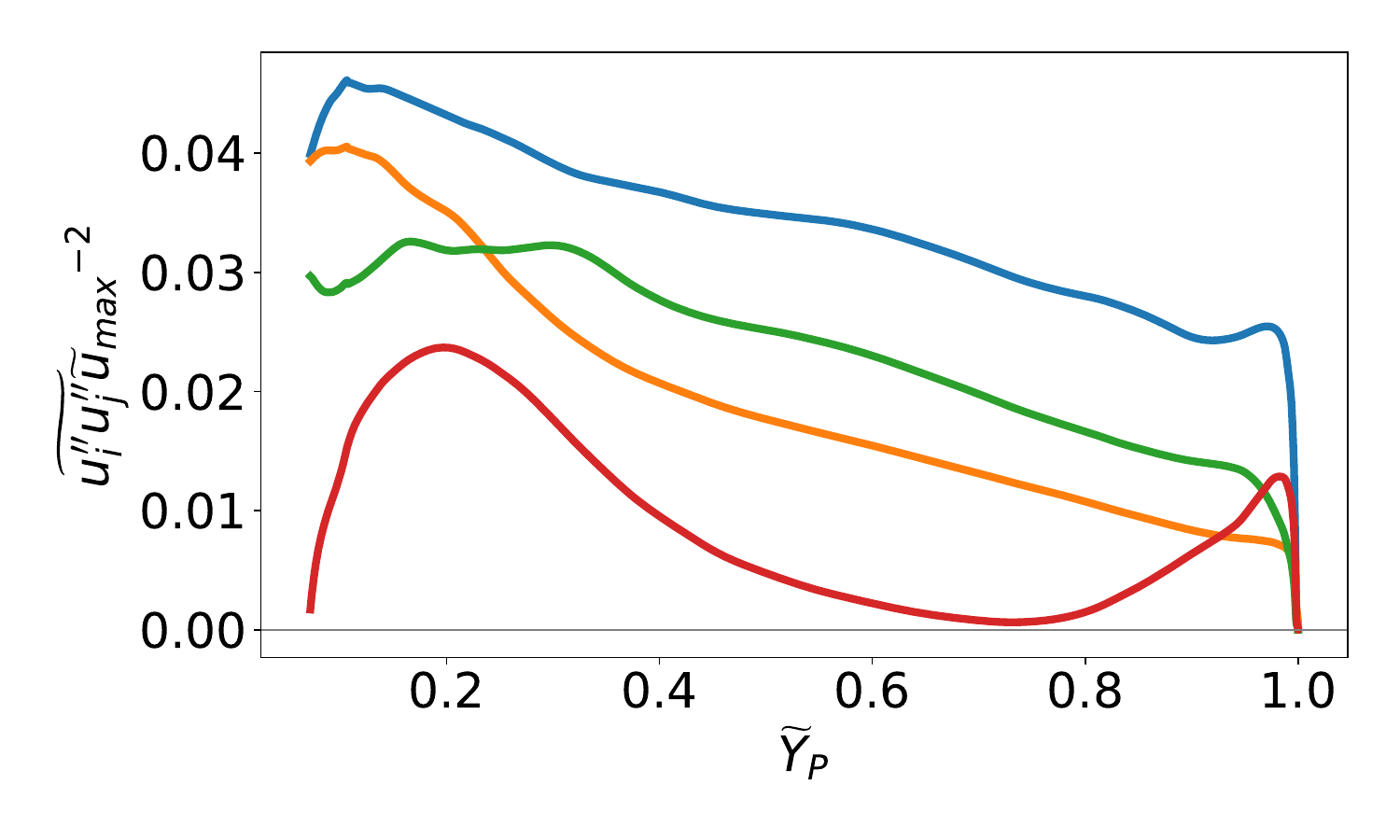} \\
    \includegraphics[width=0.49\linewidth]{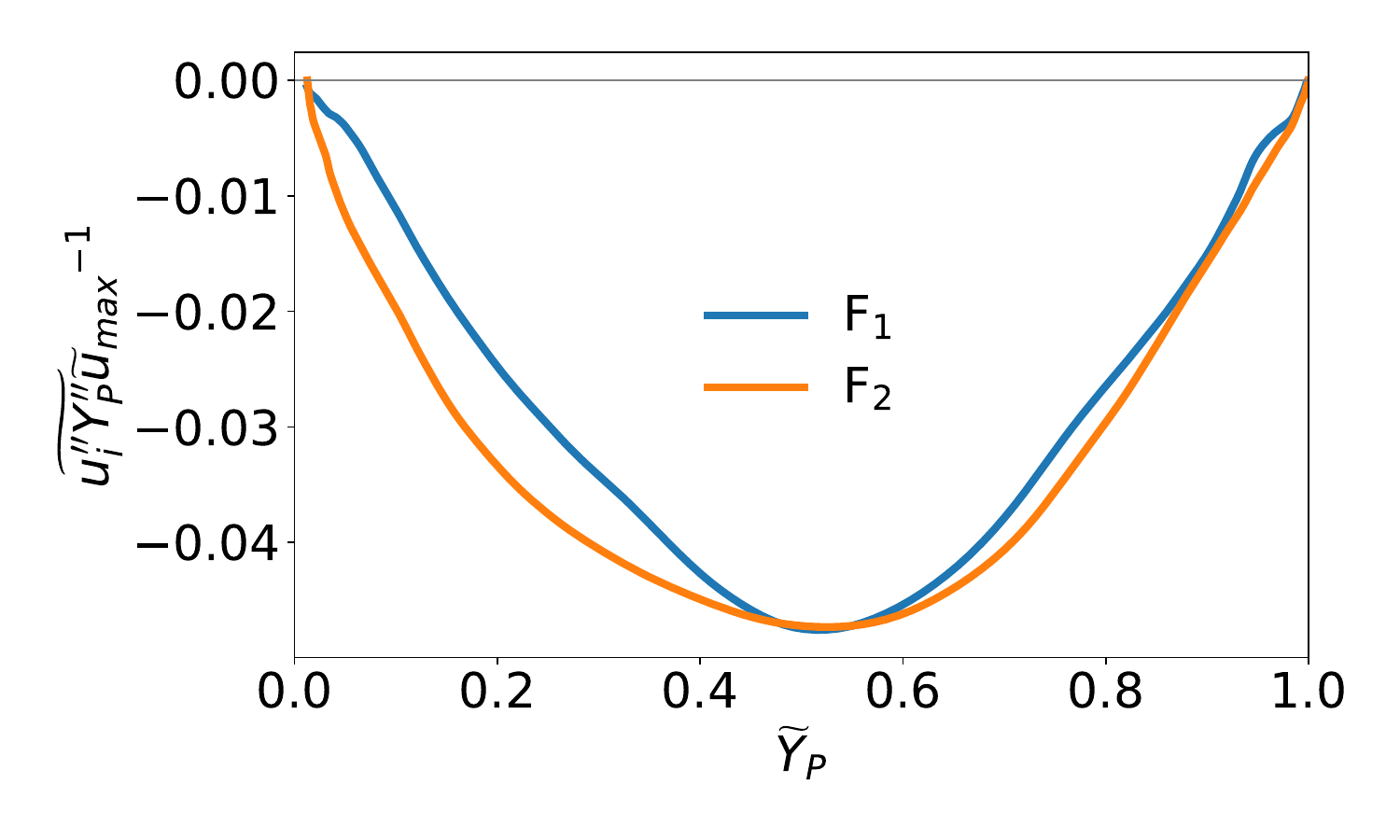}
    \hfill 
    \includegraphics[width=0.49\linewidth]{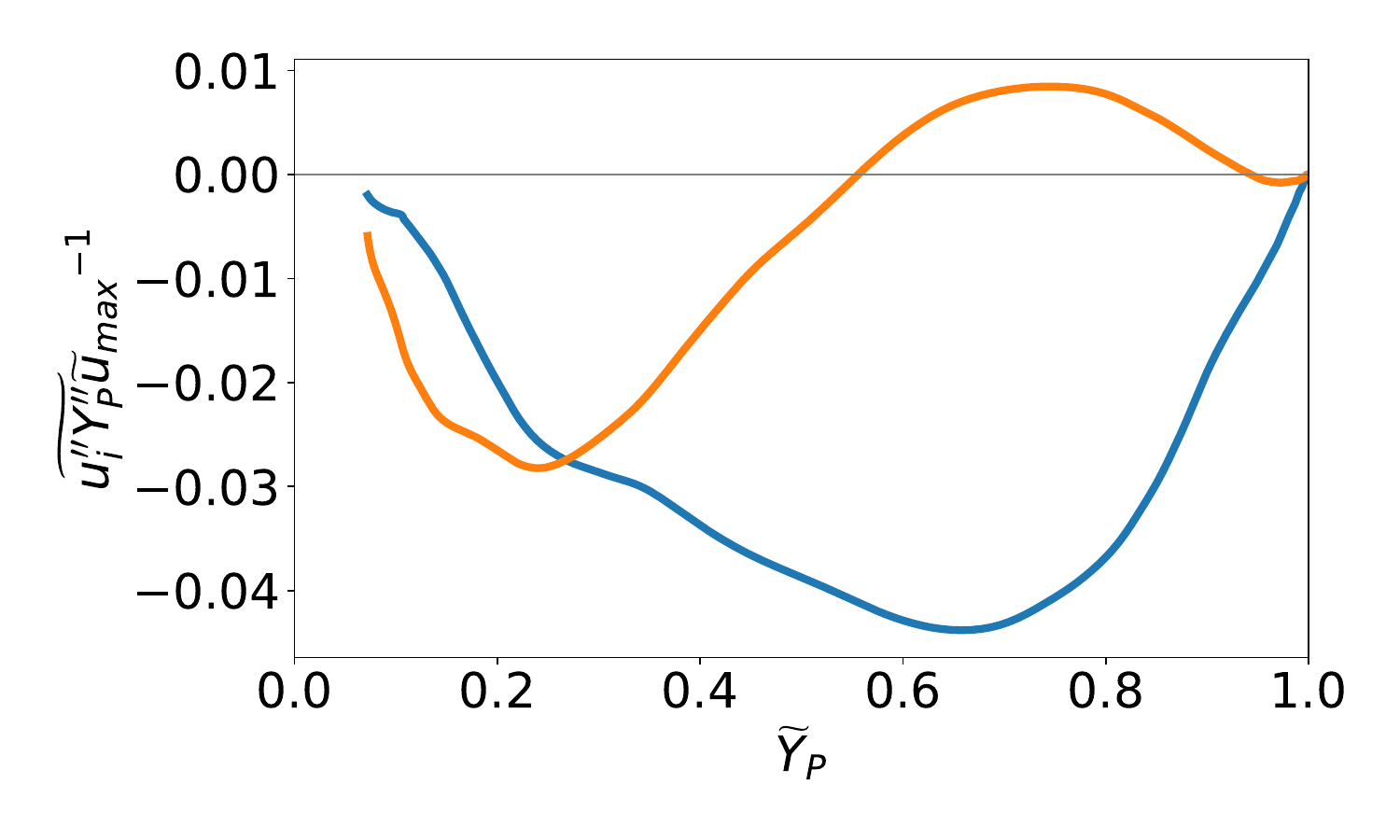}
    \caption{\emph{A priori} Reynolds stress components (top) and product species mass fraction flux components (bottom) for $\Da_{s}= 0$ (left) and $\Da_{s}= 35{,}000$ (right) at $t=32$.}
    \label{fig:DNS_reynolds_stress}
\end{figure}

We evaluate the degree of flame--turbulence interaction in the DNS datasets by comparing the \emph{a priori} (DNS-evaluated) Reynolds stress $R_{ij} = \widetilde{u_{i}^{\prime\prime}u_{j}^{\prime\prime}} = \widetilde{u_iu_j} - \wt u_i \wt u_j$ and scalar flux  $F_i = \widetilde{u_{i}^{\prime\prime}Y_{P}^{\prime\prime}} = \wt{u_i Y_P} - \wt u_i \wt Y_P$, both of which require closure in RANS. Figure~\ref{fig:DNS_reynolds_stress} plots the DNS-evaluated Reynolds stress and scalar flux components for the nonreacting case ($\Da_{s} = 0$) and the most intensely burning case ($\Da_{s} = 35{,}000$) at $t = 32$.
In the nonreacting case, the diagonal Reynolds stress components are relatively isotropic and decay monotonically through the shear layer, and the shear component ($R_{12}$) has positive sign. The nonzero scalar flux components are of similar magnitude and are both negative.

At $\Da_{s} = 35{,}000$, however, the cross-stream component ($R_{22}$) exhibits inflections within the flame due to flame-aligned pressure-dilatation \cite{macart2018effects}, and the shear component ($R_{12}$) trends negative, which is indicative of counter-Boussinesq transport. Similarly, the cross-flame scalar flux $F_2$ trends positive through the mean flame brush, which is characteristic of counter-gradient transport. These trends are consistent with the findings of MacArt \emph{et al.}~\cite{macart2018effects} for  DNS of spatially evolving turbulent premixed jet flames using detailed hydrogen--air kinetics and species transport. These trends illustrate the ability of the present one-step kinetics to recover the important flame--turbulence interactions across the Damk\"ohler number range that we assess.

\subsection{RANS equations}~\label{subsec:RANS_formulation}
We consider the popular unsteady $k$--$\epsilon$ model~\cite{launder1983numerical} as our baseline RANS closure. Its formulation includes the dimensionless turbulent kinetic energy, $k = R_{ii}/2$, which in RANS calculations is initialized from the DNS data, and the TKE dissipation rate, which is initialized as $\epsilon = 3C_\mu^{3 / 4} {k^{3 / 2}}/{{H}}$,
with $C_\mu = 0.09$. The RANS total energy is constructed as
$\wt E = \wt e + \wt u_i \wt u_i/2 + k$.

Let  $Q = [\ol{\rho}, \ol{\rho}\widetilde{u}, \ol{\rho}\widetilde{v}, \ol{\rho}\widetilde{E}, \ol{\rho}\widetilde{Y}_{P}, \ol{\rho} k, \ol{\rho}\epsilon]$ be the RANS dependent variables. We write the RANS PDEs in  implicit form as $\bh(Q, \dot Q; \theta)=0$, where $\dot{Q}=\partial Q/\partial t$, and $\theta\in\mathbb{R}^{N_\theta}$ are the tunable parameters of the neural network closure model.
With unclosed terms modeled using standard Boussinesq and gradient-diffusion closures, denoted by enumerated subscripts ``S,'' the elements of $\bh$ are the mean continuity equation
\begin{align}
    h_{\bar\rho} = \pp{\bar{\rho}}{t} + \pp{\bar{\rho}\widetilde{v}}{y} = 0,\label{eq:rans_cont}
\end{align}
the streamwise  mean momentum equation
\begin{align}
    h_{\bar\rho \tilde u} = \frac{\partial (\bar{\rho}\widetilde{u})}{\partial t} + \frac{\partial (\bar{\rho} \widetilde{u} \widetilde{v})}{\partial y} - \frac{1}{\Rey}\frac{\partial}{\partial y}\left(\mu\frac{\partial \widetilde{u}}{\partial y}\right) - \underbrace{\frac{1}{\Rey}\frac{\partial}{\partial y}\left(\mu_t(Q;\theta)\frac{\partial \widetilde{u}}{\partial y}\right)}_\mathrm{S1} = 0,\label{eq:rans_u}
\end{align}
the cross-stream mean momentum equation
\begin{align}
&h_{\bar\rho\tilde v} = \pp{\bar{\rho}\widetilde{v}}{t} + \pp{\bar{\rho} \widetilde{v} \widetilde{v}}{y} + \frac{1}{\Ma^2}\pp{\bar{p}}{y} - \frac{1}{\Rey}\frac{4}{3}\pp{}{y}\left(\mu\pp{\widetilde{v}}{y}\right) - \underbrace{\frac{1}{\Rey}\frac{4}{3}\pp{}{y}\left(\mu_t(Q;\theta)\pp{\widetilde{v}}{y}\right) + \pp{}{y}\left(\frac{2}{3}\bar{\rho}k\right)}_\mathrm{S2} = 0,\label{eq:rans_v}
\end{align}
the mean total energy equation
\begin{align}
  h_{\bar\rho\tilde E} &= \frac{\partial (\bar{\rho} \widetilde{E})}{\partial t}  
  + \frac{\partial}{\partial y} \left( \widetilde{v} \bar{\rho} \widetilde{E} \right)
  + \frac{1}{\Ma^2}\frac{\partial (\bar{p} \widetilde{v})}{\partial y}
  - \frac{1}{\Ma^2 \Rey {\Pr} } \frac{\partial}{\partial {y}}\left({{{\mu}}} \frac{\partial \widetilde{{T}}}{\partial {y}}\right)
  - \underbrace{{\frac{1}{\Ma^2 \Rey {\Pr_{t}}} \frac{\partial}{\partial {y}}\left({{\mu_{t,E}(Q;\theta)}} \frac{\partial \widetilde{{T}}}{\partial {y}}\right)}}_\mathrm{S3}  \nonumber \\
  &- \frac{1}{\Rey } \frac{\partial}{\partial {y}} \left( \widetilde{{u}}\mu\frac{\partial {\widetilde{u}}}{\partial {y}} + \frac{4}{3}\widetilde{{v}}\mu\frac{\partial {\widetilde{v}}}{\partial {y}} \right)
  - \underbrace{\frac{1}{\Rey}\frac{\partial}{\partial y} \left( \left(\mu + \frac{\mu_t(Q;\theta)}{\sigma_k}\right) \frac{\partial k}{\partial y} \right)}_\mathrm{S4} 
  - \frac{1}{\Ma^2 \Rey \Sc \gamma} \frac{\partial}{\partial {y}} \left( \rho \mu \sum_{k=1}^{N} \widetilde{h}_k\pp{\widetilde{Y}_{k}}{y} \right) \nonumber \\
  &+ \underbrace{\frac{C_{h3}}{\Ma^2\Rey\Sc_{t}\gamma}{\frac{\partial}{\partial y}\left({\mu_t(Q;\theta)}\frac{\partial \widetilde{Y}_{P}}{\partial {y}}\right)}}_\mathrm{S5} 
  - \underbrace{ \frac{1}{\Rey }\frac{\partial}{\partial {y}} \left( \widetilde{{u}}\mu_t(Q;\theta)\frac{\partial {\widetilde{u}}}{\partial {y}} + \frac{4}{3}\widetilde{{v}}\mu_t(Q;\theta)\frac{\partial {\widetilde{v}}}{\partial {y}} \right) + \pp{}{y}\left(\frac{2}{3}\widetilde{{v}}\bar{{\rho}}{k}\right)}_\mathrm{S6} = 0, \label{eq:rans_E}
\end{align}
the mean product species equation 
\begin{align}
&h_{\bar\rho\tilde Y_{P}} = \frac{\partial \bar{\rho} \widetilde{Y}_{P}}{\partial t} + \frac{\partial}{\partial y}\left(\bar{\rho} \widetilde{Y}_{P} \widetilde{v}\right) - \frac{1}{\Rey \Pr{\Le}}\frac{\partial}{\partial y}\left({\rho \mu}\pp{\widetilde{Y}_{P}}{y}\right) - \underbrace{\frac{1}{\Rey\Sc_{t}}\frac{\partial}{\partial y}\left({\mu_t(Q;\theta)}\pp{\widetilde{Y}_{P}}{y}\right)}_\mathrm{S7} - \Da_{s} W {\dot{\omega}_{{P}}{(\overline{\rho},\widetilde{Y_{P}},\widetilde{T})}} = 0,\label{eq:rans_Y}
\end{align}
and the \( k \) and \( \epsilon \) transport equations 
\begin{align}
    h_{\bar\rho k} &= \frac{\partial (\bar{\rho} k)}{\partial t} + \frac{\partial \left(\bar{\rho} k \widetilde{v}\right)}{\partial y} - \frac{1}{\Rey}\frac{\partial}{\partial y}\left[\left(\mu + \frac{\mu_t(Q;\theta)}{\sigma_k}\right)\frac{\partial k}{\partial y}\right] - \frac{1}{\Rey}\mu_t(Q;\theta)\left[\left(\pp{\widetilde{u}}{y}\right)^{2} +2\left(\pp{\widetilde{v}}{y}\right)^{2}\right] + \bar{\rho} \varepsilon = 0,\label{eq:rans_k}\\
    h_{\bar\rho \epsilon} &= \frac{\partial (\bar{\rho} \varepsilon)}{\partial t} + \frac{\partial \left(\bar{\rho} \varepsilon \widetilde{v}\right)}{\partial y} - \frac{1}{\Rey}\frac{\partial}{\partial y}\left[\left(\mu + \frac{\mu_t(Q;\theta)}{\sigma_\epsilon}\right)\frac{\partial \varepsilon}{\partial y}\right] - \frac{1}{\Rey}C_{1 \varepsilon}\mu_t(Q;\theta)\frac{\varepsilon}{k}\left[\left(\pp{\widetilde{u}}{y}\right)^{2} + {2}\left(\pp{\widetilde{v}}{y}\right)^{2}\right]  \nonumber \\
    &+ C_{2 \varepsilon}\bar{\rho}\frac{\varepsilon^2}{k} = 0.\label{eq:rans_epsilon}
\end{align}
In these, $\mu_t(Q;\theta)$ and $\mu_{t,E}(Q;\theta)$ are modeled eddy viscosities (presented in Section~\ref{sec:eddyvisc}), $\Pr_t=0.9$ is the constant turbulent Prandtl number, $\Sc_t = 0.65$ is the constant turbulent Schmidt number, and \( \sigma_k = 1.00 \), \( \sigma_{\varepsilon} = 1.30 \), \( C_{1 \varepsilon} = 1.44 \), and \( C_{2 \varepsilon} = 1.92 \) are  standard \keps\ model constants.
The full derivation of \eqref{eq:rans_E}, including the species enthalpy flux constant $C_{h3} = 17.422$, is provided in Appendix A.
In \eqref{eq:rans_Y}, the chemical source term is evaluated directly using the averaged variables, which introduces an additional unclosed term due to the noncommutativity of averaging with the nonlinear source term. This additional error is neglected in order to focus on the turbulent transport terms. The impact of this is assessed in Section~\ref{subsec:in-sample}.

Solving \eqref{eq:rans_cont}--\eqref{eq:rans_epsilon} on the DNS mesh ($N_{y} = 800$) would require time steps of size $\Delta_{t} = 7.5 \times 10^{-4}$ to ensure numerical stability ($\mathrm{CFL} \leq 0.8)$. This would  require 25,200 time steps per optimization epoch (see Section~\ref{sec:DPM}) and would make model optimization prohibitively expensive. Since RANS does not resolve the turbulence scales, a coarser RANS mesh of $N'_{y} = 512$ with stretching ratio $\delta'_{y} = 3.0$ is employed, which allows for larger time steps of size $2.7 \times 10^{-3}$ and reduces the number of time steps per simulation to $N_T=7{,}000$. This coarser mesh retains adequate resolution of the mean flame and shear layer thicknesses. A mesh independence study confirms no significant differences in the RANS solution fields between the two meshes. 
\subsection{Eddy-viscosity closures} \label{sec:eddyvisc}

The eddy-viscosity terms in \eqref{eq:rans_cont}--\eqref{eq:rans_epsilon} are our primary targets for neural network augmentation. For the baseline \keps\ model, the eddy viscosity for turbulent momentum transport is given by
\begin{equation}
  \label{eq:rans_mut_kep}
  \mu_t(Q) = \mu_t^{k\epsilon}(\{k,\epsilon\};C_{\mu}) = \Rey \, C_{\mu} \, \bar{\rho} \frac{k^{2}}{\epsilon},
\end{equation}
and the eddy viscosity used in the turbulent heat flux is $\mu_{t,E}=\mu_t$, which is converted to a turbulent thermal conductivity via the constant Prandtl number.

The method proposed for a generalized combustion RANS model involves assimilating data into the modeled PDEs using neural networks. This approach builds upon previous DPM approaches for nonreacting LES and RANS closures~\cite{sirignano2020dpm,macart2021embedded,sirignano2023deep,sirignano2023pde}. We embed an untrained neural network $\mathcal{N}$ into the baseline RANS framework to augment the eddy viscosities as
\begin{equation}
  \label{eq:NN formulation}
  \mu_t(Q;\theta), \mu_{t,E}(Q;\theta)
  \leftarrow \mathcal{N}\Biggl( \frac{\partial \bar{\rho}}{\partial y},\, \frac{\partial \bar{\rho} \widetilde{u}}{\partial y},\, \frac{\partial \bar{\rho}\widetilde{v}}{\partial y},\, \frac{\partial \bar{\rho}\widetilde{E}}{\partial y},\, \frac{\partial \bar{\rho} \widetilde{Y}_P}{\partial y},\, \frac{\partial \bar{p}}{\partial y},\, \mu_t^{k\epsilon},\, \Da_s\, {\dot{\omega}_{{P}}{(\overline{\rho},\widetilde{Y_{P}},\widetilde{T})}},\, t \Biggr),
\end{equation}
where $\mathcal{N}$ accepts inputs as the flow state gradients at each cell (and its two nearest neighbors), the local $\mu_t^{k\epsilon}$ from \eqref{eq:rans_mut_kep}, the local RANS chemical source term, and the simulation time.

In parametric studies of neural network outputs including (i) an additional turbulent viscosity in the species equation \eqref{eq:rans_Y} and (ii) directly modeling Reynolds stress and scalar flux, using $\mu_t(Q;\theta)$ and $\mu_{t,E}(Q;\theta)$ not only achieves comparable or improved accuracy to these but also enhances out-of-sample robustness. In contrast, directly modeling the unclosed terms (here, the Reynolds stress and scalar flux) with a neural network is known to introduce instabilities during training~\cite{wu2019reynolds}.  
Furthermore, the inclusion of time ($t$) as an input to the neural network is specific to the present time-evolving RANS predictions, in which the network aims to recover particular ensemble-averaged trajectories. For a statistically stationary flow, the input vector in \eqref{eq:NN formulation} would be extended spatially, and the time coordinate would be omitted.

The neural network is implemented as a multilayer perceptron (MLP) comprising four layers with 300 nodes each, totaling 189,000 trainable parameters. Each hidden layer employs rectified linear unit (ReLU) activation functions, while the final layer uses exponential linear unit (ELU) activation functions~\cite{clevert2015fast}. The network is evaluated pointwise on the computational mesh (i.e., it receives only local data), with symmetry enforced by evaluating the network on one half of the mesh and mirroring its output across the centerline. This approach ensures that the model satisfies the required symmetry of the jet statistics by narrowing the search space to functions that adhere to the established physical symmetries, thereby reducing the risk of unphysical predictions.

\section{Adjoint-based PDE augmentation}
\label{sec:DPM}
We train the PDE-embedded neural networks using adjoint-based optimization. The PDE-constrained optimization problem to be solved is 
\begin{equation}
  \argmin_\theta \bar J(Q(\theta))\quad \text{subject to}\quad \bh(Q,\dot Q;\theta)=0,
  \label{eq:optim_problem}
\end{equation}
that is, to find the parameters that minimize the objective (loss) function $\bar J$ while satisfying \eqref{eq:rans_cont}--\eqref{eq:rans_epsilon}. In \eqref{eq:optim_problem}, we emphasize the implicit dependence of $\bar J$ on the parameters via the RANS PDE solution $Q$. We use a time-integrated objective function
\begin{equation}
    \label{eq:time_integrated_loss}
    \bar J(Q(\theta)) = \sum_{n=1}^{N_\mathrm{op}} J(Q^n(\theta)) \Delta t,
\end{equation}
in which the hyperparameter $N_\mathrm{op}$ specifies the number of simulation time steps per optimization window; hence, each optimization window spans a time horizon $\tau_\mathrm{op}=N_\mathrm{op}\Delta t$. One optimization window comprises a forward pass of $N_\mathrm{op}$ steps, an adjoint pass of $N_\mathrm{op}$ steps, and one gradient-descent parameter update.

The choice of $\tau_\mathrm{op}$ depends on the physical time scales of the system. Optimization windows of too small \( \tau_{\mathrm{op}} \) can provide insufficient observations of the system's dynamics, thus limiting the network’s ability to augment the unsteady PDEs, while too large \( \tau_{\mathrm{op}} \) can cause excessive adjoint magnitudes relative to the Jacobian determinant, leading to machine-precision roundoff errors \cite{liu2024adjoint}. We use $N_\mathrm{op}=100$ ($\tau_\mathrm{op}=0.27$) for all trained models due to its balance of optimization window duration and manageable adjoint magnitudes.
In terms of the  centerline ($y=0$) integral time scale $\tau_\mathrm{\ell}=\ell/u'$, where the integral length scale $\ell$ is evaluated from the two-point streamwise velocity autocorrelation at $t=25$, this corresponds to $\tau_\mathrm{\ell}/\tau_\mathrm{op}$ between 11.41 ($\Das=0$) and 21.04 ($\Das=13{,}000$).

In \eqref{eq:time_integrated_loss}, the instantaneous objective function is
\begin{equation}
\label{eq:J loss}
J(Q^n(\theta)) = \left[ S_c \odot \left( Q^n(\theta) - Q^{e,n} \right)^2 \right]^{\top} \mathbf{1},
\end{equation}
in which $Q^{e,n}\in\mathbb{R}^7$ are target fields at the same time instant obtained from ensemble-averaged DNS data, $\odot$ represents elementwise (Hadamard) multiplication, and $\mathbf{1}\in\mathbb{R}^7$ is a vector of ones.
Constant scaling factors $S_{c} = [1.138, 2.5, 29.06, 0.015, 8.33, 0, 0]$ are applied to balance the relative magnitudes of the terms in \eqref{eq:J loss}. This balance facilitates effective learning across the dependent variables. The scaling factors are obtained as the inverse of the time-averaged maximum values of each conserved variable over the full RANS predictions across all $\Da_{s}$. The scaling factors for $\ol{\rho} k$ and $\ol{\rho} \epsilon$ are zero to eliminate the influence of these variables in \eqref{eq:J loss}.

The gradient $\nabla_{\theta} \bar J$ needed for gradient-descent optimization could be evaluated as
\begin{equation}
\label{eq:Jtheta}
    \nabla_{\theta} \bar J(Q(\mathbf{\theta})) = \frac{\partial \bar J}{\partial  Q} \frac{\partial Q}{\partial \mathbf{\theta}},
\end{equation}
for example, using perturbations $\theta'=\theta + \Delta\theta$ to calculate the second factor.
However, this would be prohibitively expensive for large $N_\theta$, as is common in deep learning, due to the implicit dependence of $Q$ on $\theta$ via the PDE---i.e., the PDE solution would need to be recalculated for each perturbation.
Instead, the PDE constraint may be introduced using the Lagrange multipliers (adjoint variables) $\check\chi=[\check\Omega_{\bar{\rho}}, \check{\Omega}_{\bar{\rho} \tilde{u}},\check{\Omega}_{ \bar{\rho} \tilde{v}},\check{\Omega}_{\bar{\rho} \tilde{Y}_{p}},\check{\Omega}_{\bar{\rho} \tilde{E}},\check{\Omega}_{\bar{\rho} k}, \check{\Omega}_{\bar{\rho} \epsilon}]$ 
to construct a Lagrangian
\begin{align}
  \label{eq:Lagrangian}
  \mathcal{L} &\equiv \bar J + \check\chi^\top \bh,
\end{align}
which is equivalent to $\bar J$ when the PDE constraint $\bh(Q,\dot Q; \theta)=0$ is satisfied. Differentiating \eqref{eq:Lagrangian} with respect to $\theta$, expanding using the chain rule, and ensuring that terms multiplying $\partial Q/\partial\theta$ reduce to zero requires $\check\chi$ to satisfy 
\begin{equation}
  \label{eq:u_hat_ode}
  \frac{d {\check{\chi}}}{d t}=\check\chi^{\top} \frac{\partial {{\bh}}}{\partial  Q}+\frac{\partial J}{\partial  {Q}}
\end{equation}
over $t\in[0, N_\mathrm{op}]\Delta t$ with ``initial'' condition $\check\chi(N_\mathrm{op}\Delta t) = 0$. In \eqref{eq:u_hat_ode}, the partial derivative $\partial J / \partial  Q$ is calculated analytically, and the term ${\check{\chi}}^{\top} \partial {\bh} / \partial  Q$ is calculated using algorithmic differentiation over the RANS PDEs.
After solving \eqref{eq:u_hat_ode} in reverse time (the ``backward pass''), the gradient needed for optimization is obtained using $\check\chi$ as
 \begin{align}
 \label{eq:grad_L}
 \nabla_\theta \bar{J}=\nabla_\theta \mathcal{L}=\sum_{n=1}^{N_\mathrm{op}} \check{{\chi}}_{n}^{\top} \left.{\frac{\partial {\bh}}{\partial \theta}}\right|_{n} \Delta t.
 \end{align}

\begin{figure}
    \centering
    \includegraphics[width=1.0\linewidth]{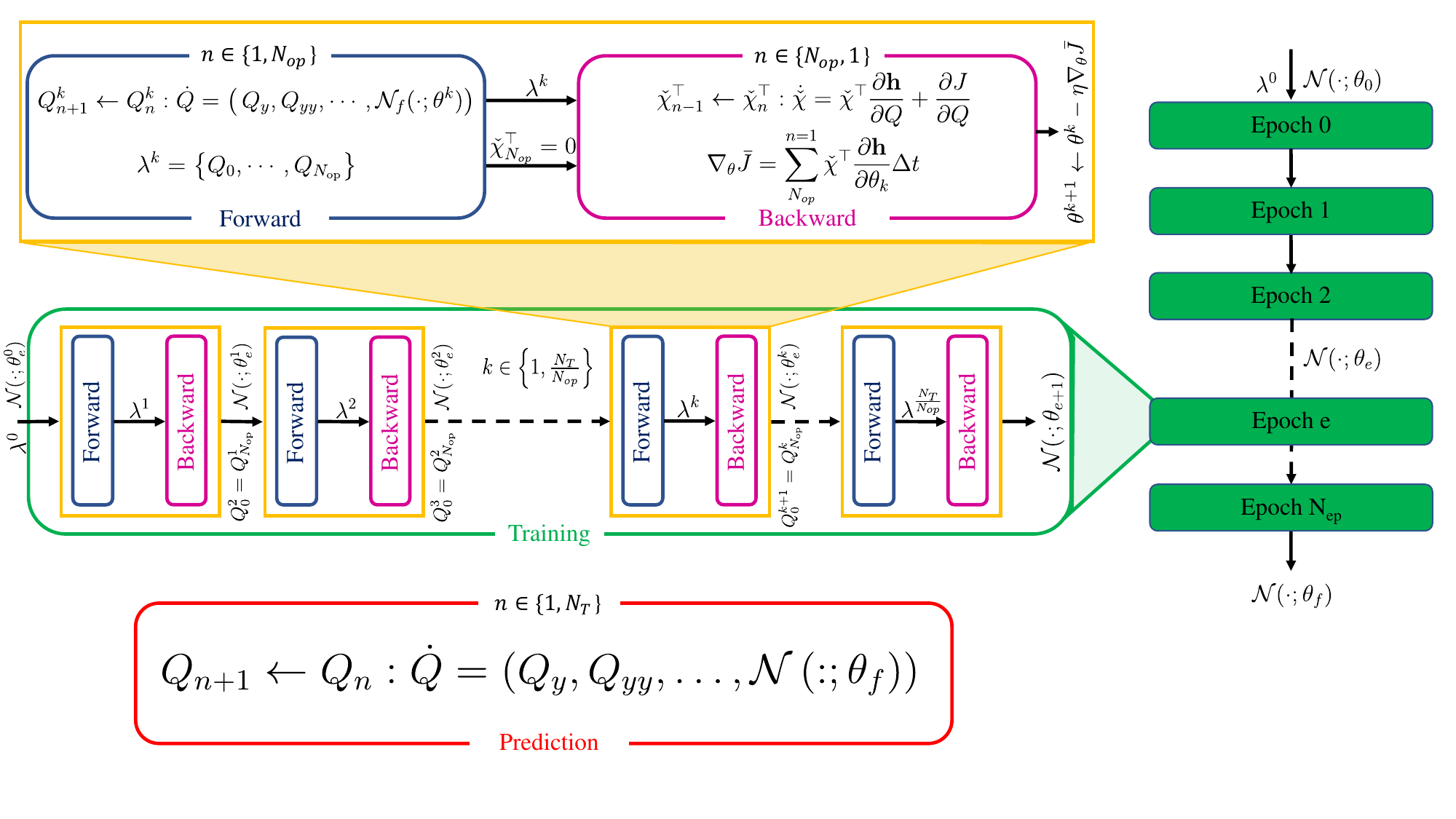}
    \caption{Schematic of unsteady, PDE-constrained, adjoint-based optimization framework for PDE-embedded neural networks.}
    \label{fig:alg}
\end{figure}

The solution of \eqref{eq:optim_problem} is divided into ``epochs'' of \(N_{T}=7000\) time steps. Each epoch corresponds to a full forward RANS solution and the adjoint backward steps to train the network. Consequently, each epoch comprises $N_T/N_\mathrm{op}$ optimization iterations. 
 This algorithm is depicted in Figure~\ref{fig:alg}, showing the use of \eqref{eq:u_hat_ode} and \eqref{eq:grad_L} in the backward (adjoint) passes. The forward pass updates the conserved variables \(Q\) over time, while the backward pass updates the adjoint variables \({\check{\chi}}\) utilizing checkpointed states \(\lambda^k\). The neural network parameters \(\theta\) are updated after each backward pass.  Multiple epochs are performed until \eqref{eq:time_integrated_loss} converges. 
 
\begin{figure}
    \centering
    \includegraphics[width=0.6\linewidth]{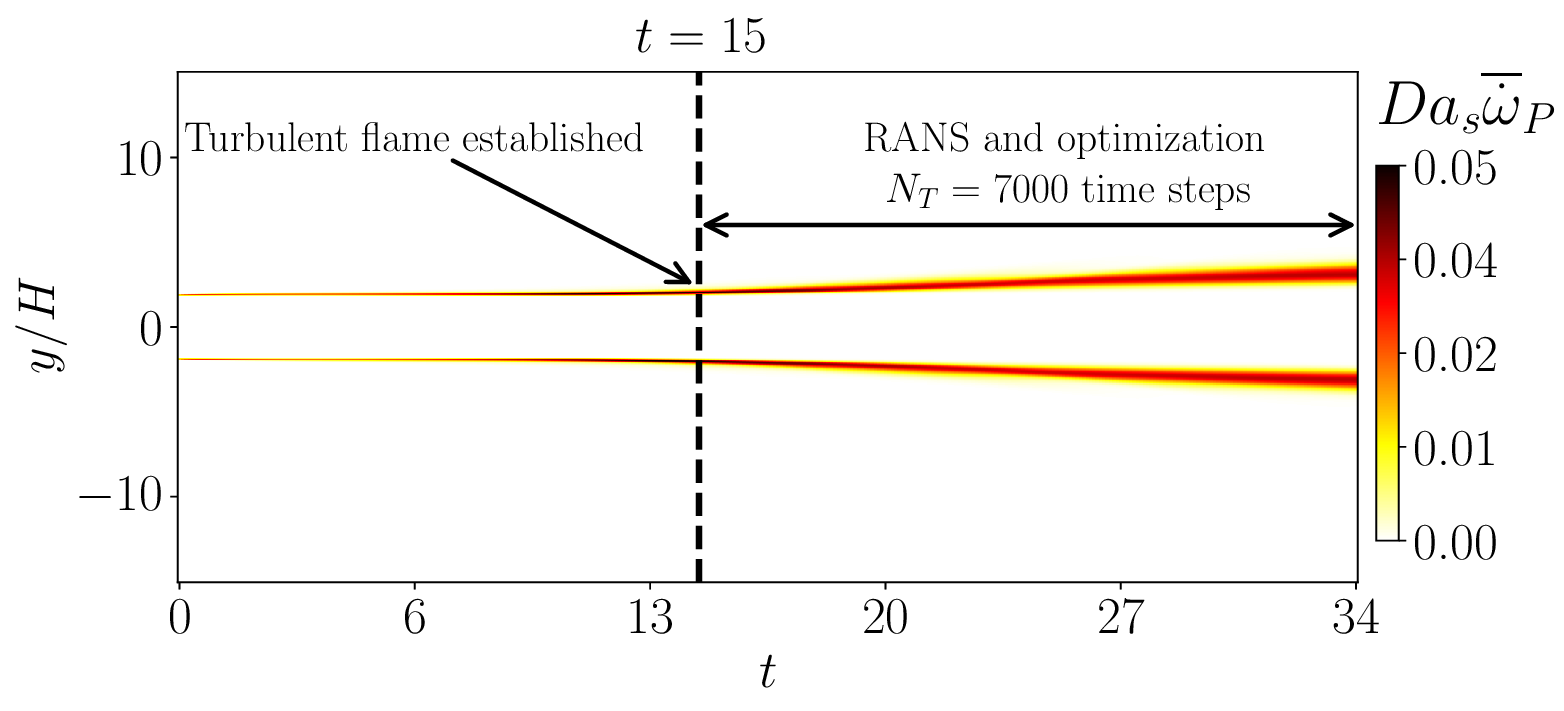}
    \caption{Turbulent flame establishment at $t=15$ and epoch length $N_T$. Shown for the \emph{a priori} (DNS-evaluated) averaged heat release rate  for $\Da_{s} = 20{,}000$.}
    \label{fig:flame_time_scale}
\end{figure}

Figure~\ref{fig:flame_time_scale} illustrates the $N_T=7000$  epoch length, which corresponds to 18.9 time units ($t=15$ to $t=33.9$) after the establishment of the turbulent flame at $t=15$. Notably, the establishment of a self-similar turbulent jet prior to \(t=15\) is nearly identical across the scaling Damk\"ohler numbers.  Optimizing over this initial flow evolution would offer marginal additional insight while extending the training time.


Compared to other PDE-coupled optimization methods that require continuous tracking of the computational graph over time (e.g., \cite{akhare2023physics}), the adjoint-based approach is computationally advantageous for unsteady problems because it breaks the computational graph between each time step of the backward pass. This significantly reduces the memory burden encountered in non-graph-breaking approaches, which suffer increasing memory requirements with increasing $N_\mathrm{op}$ (for example, methods that rely upon algorithmic differentiation over the entire backward pass). Instead, the present method propagates sensitivities over the backward pass using the adjoint variables.
By breaking the computational graph using \eqref{eq:grad_L}, the adjoint method's memory usage remains constant over the backward pass for any $N_\mathrm{op}$. While the method requires solving additional differential equations \eqref{eq:u_hat_ode}, this additional cost remains small for the size of PDE systems typically encountered in turbulence and combustion.

\subsection{Training configuration}

In our training configuration, we initialize the learning rate as \(\alpha = 10^{-3}\) and reduce it by half if \(\bar J\) does not decrease over four consecutive epochs, ensuring adaptive control as training progresses. We employ the RMSprop~\cite{hinton2012neural} optimizer---an adaptive learning rate method that addresses the diminishing learning rate problem by maintaining a moving average of the gradient magnitude \(\nabla_\theta \bar J\) to normalize the gradient-descent steps. Training proceeds for 200 epochs per model, after which significant additional convergence was not observed. Further details of the optimization convergence and sensitivity of the training process to model initialization are provided in \ref{sec_appen: training_convergence}.

The neural network model \eqref{eq:NN formulation} that predicts \(\mu_t\) and \(\mu_{t,E}\) is designated ``ME.'' Models are trained for both single and multiple scaling Damk\"ohler numbers. Single-\(\Da_s\)-trained models are designated ME\(x\), where \(x = \Da_s/10{,}000\). For example, ME0 is trained for \(\Da_s = 0\), and ME6 is trained for \(\Da_s = 6{,}000\); other models are similarly trained for different values of \(\Da_s\).

\subsection{Multiple-condition training}
To evaluate data scalability, we train a model in parallel for $\Da_s \in \{0, 13{,}000, 20{,}000, 35{,}000\}$. Each of the simultaneous predictions uses the same neural network with shared parameters. After each backward pass, the gradients obtained from each simulation are averaged using message-passing interface (MPI) communication, and the parameters are updated using these shared gradients. The adaptive learning rate is dynamically set to the lowest learning rate among the simultaneous simulations.

To accelerate convergence, new models can be initialized using the parameters of a model ``pretrained'' for a different Damk\"ohler number; these are additionally designated ``PT'' with the original training Damk\"ohler number in parentheses. For example, ME(0,13,20,35)-PT(ME35) is a model trained in parallel with initial parameters obtained from the converged ME35 model.

\begin{figure}
    \centering    
    \includegraphics[width=\linewidth]{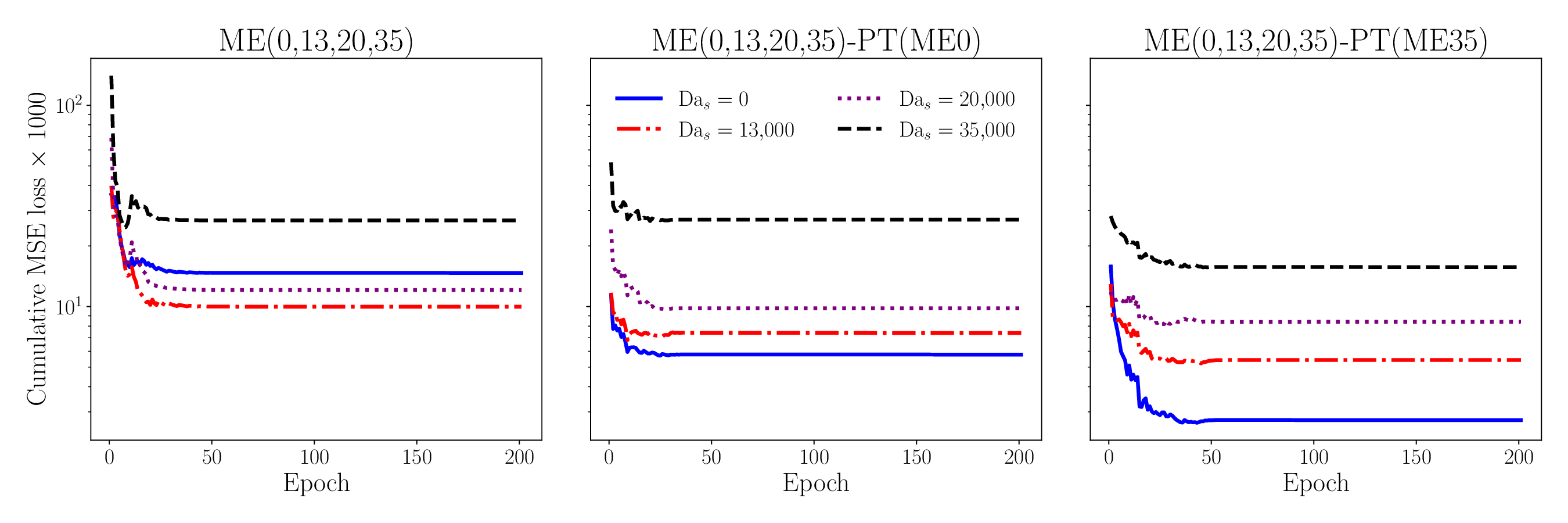}
    \caption{Training on multiple $\Da_{s}$ datasets with randomly initialized parameters (left), pretrained parameters initialized from ME0 (center), and pretrained parameters initialized from ME35 (right).}
    \label{fig:train_loss_combined}
\end{figure}

Figure~\ref{fig:train_loss_combined} shows the training loss convergence for multiple-Damk\"ohler number-trained models using random parameter initialization, initialization from the pretrained ME0 model, and initialization from the pretrained ME35 model.  Random initialization produces training losses that decrease for all $\Da_s$ values; however, this approach has higher initial training loss and requires more iterations to  converge.  Initialization using the trained ME0 model (nonreacting case) improves the initial loss compared to random initialization but struggles to converge for higher $\Da_s$ values, especially for the most intensely burning case ($\Da_s = 35{,}000$) due to the differences in dynamics between the nonreacting and reacting cases. Initialization using the ME35 model biases the pretraining to the highest Damk\"ohler number ($\Da_s = 35{,}000$), which successfully reduces training losses across all $\Da_s$ values and the $\Da_s = 35{,}000$ case in particular. The parallel-trained models' ability to generalize across the Damk\"ohler number range, as well as the impact of parameter initialization on prediction accuracy, is further explored in Section~\ref{sec:Results}.

\section{Results}
\label{sec:Results}
This section evaluates the performance of neural network-augmented RANS models for in-sample cases in Section~\ref{subsec:in-sample}.  Section~\ref{subsec:generalization} examines the generalization capabilities of the trained models for both single- and multiple-$\Da_s$ training. Section~\ref{subsec:weight_init} examines the effect of weight initialization on multiple-$\Da_s$ training.

\subsection{In-sample evaluation for $\Da_s=0$ and $\Da_s=20{,}000$}
\label{subsec:in-sample}
\begin{figure}
  \centering
  \includegraphics[width=0.49\linewidth]{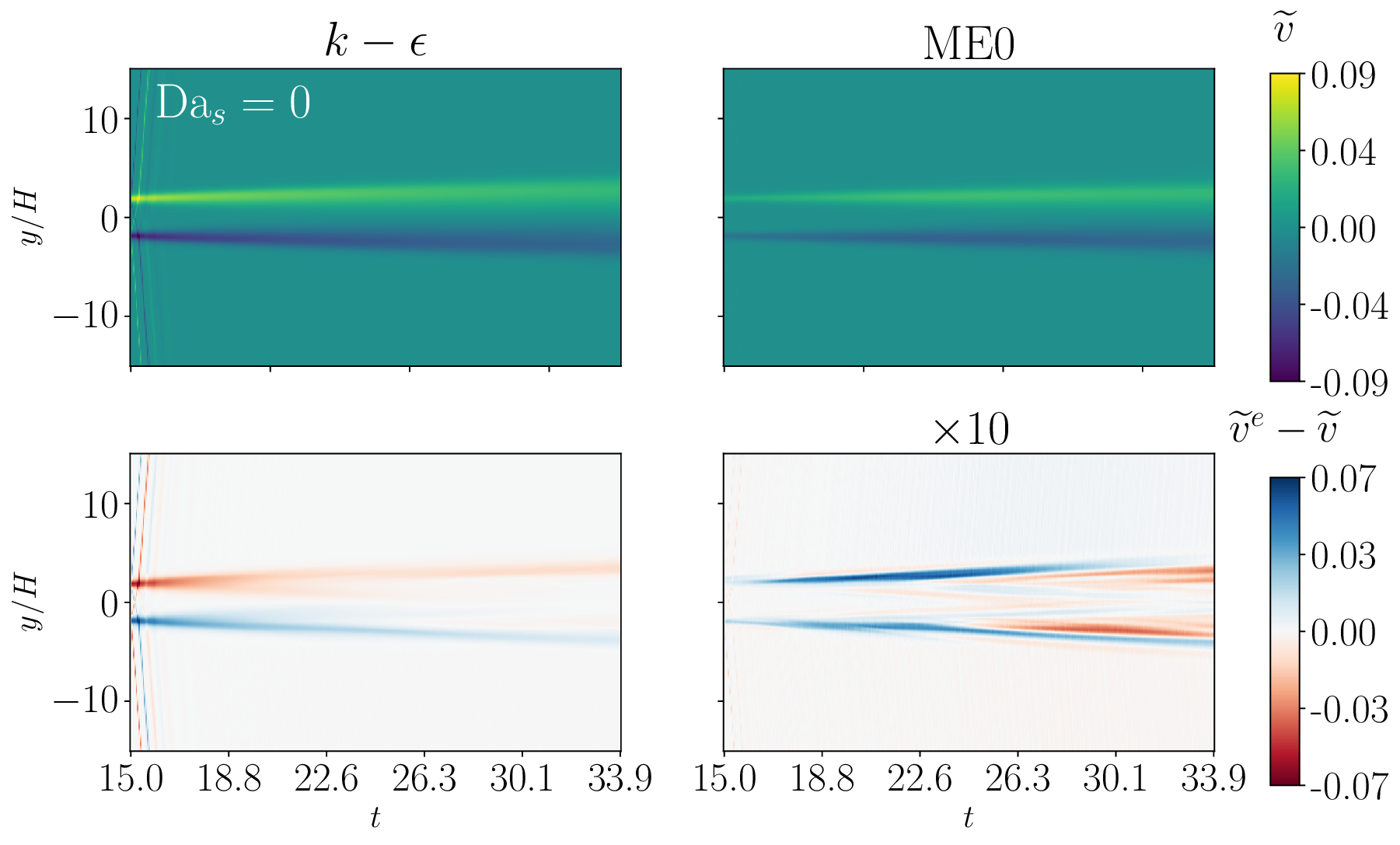}
  \hfill
  \includegraphics[width=0.49\linewidth]{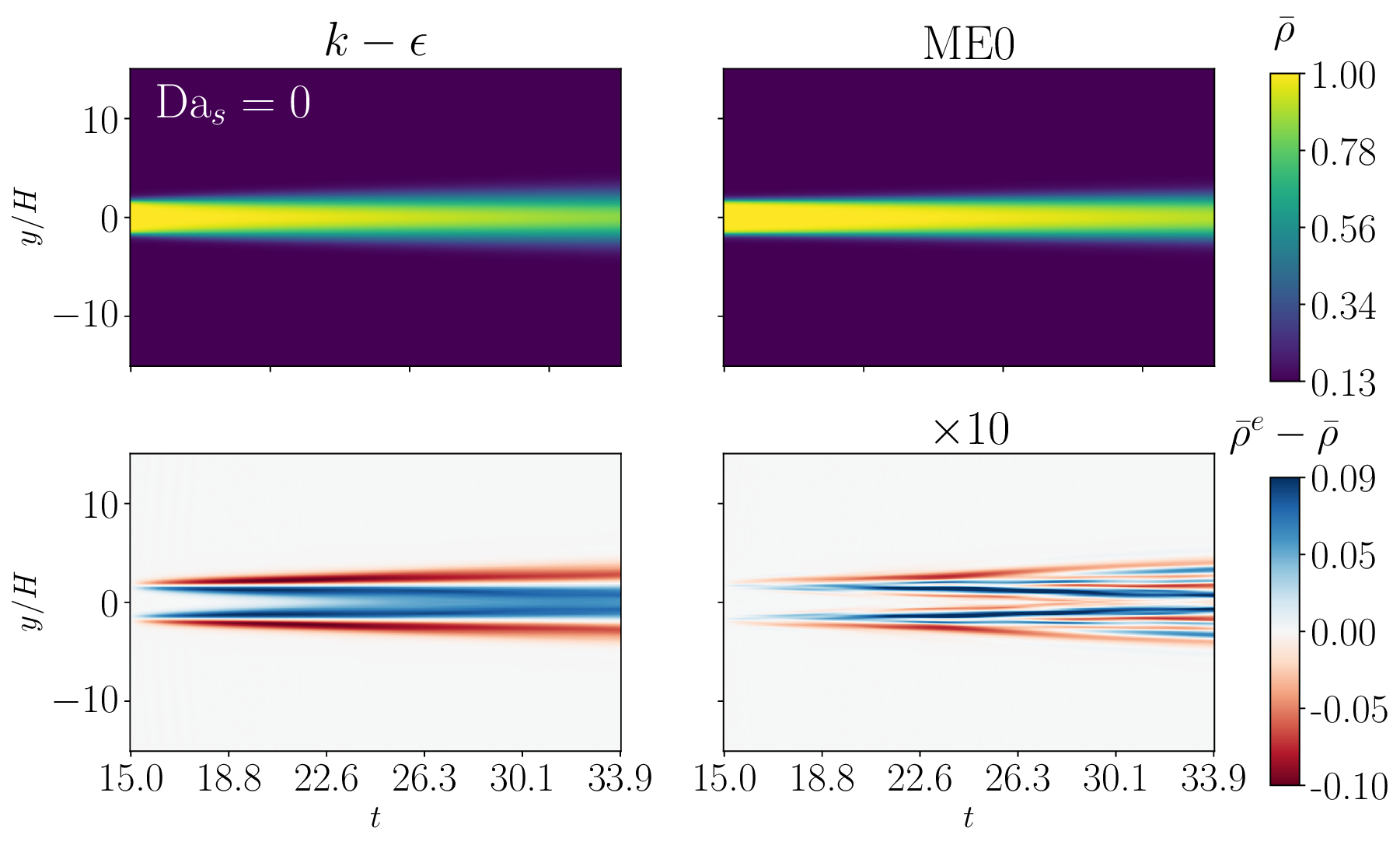} \\
  \includegraphics[width=0.49\linewidth]{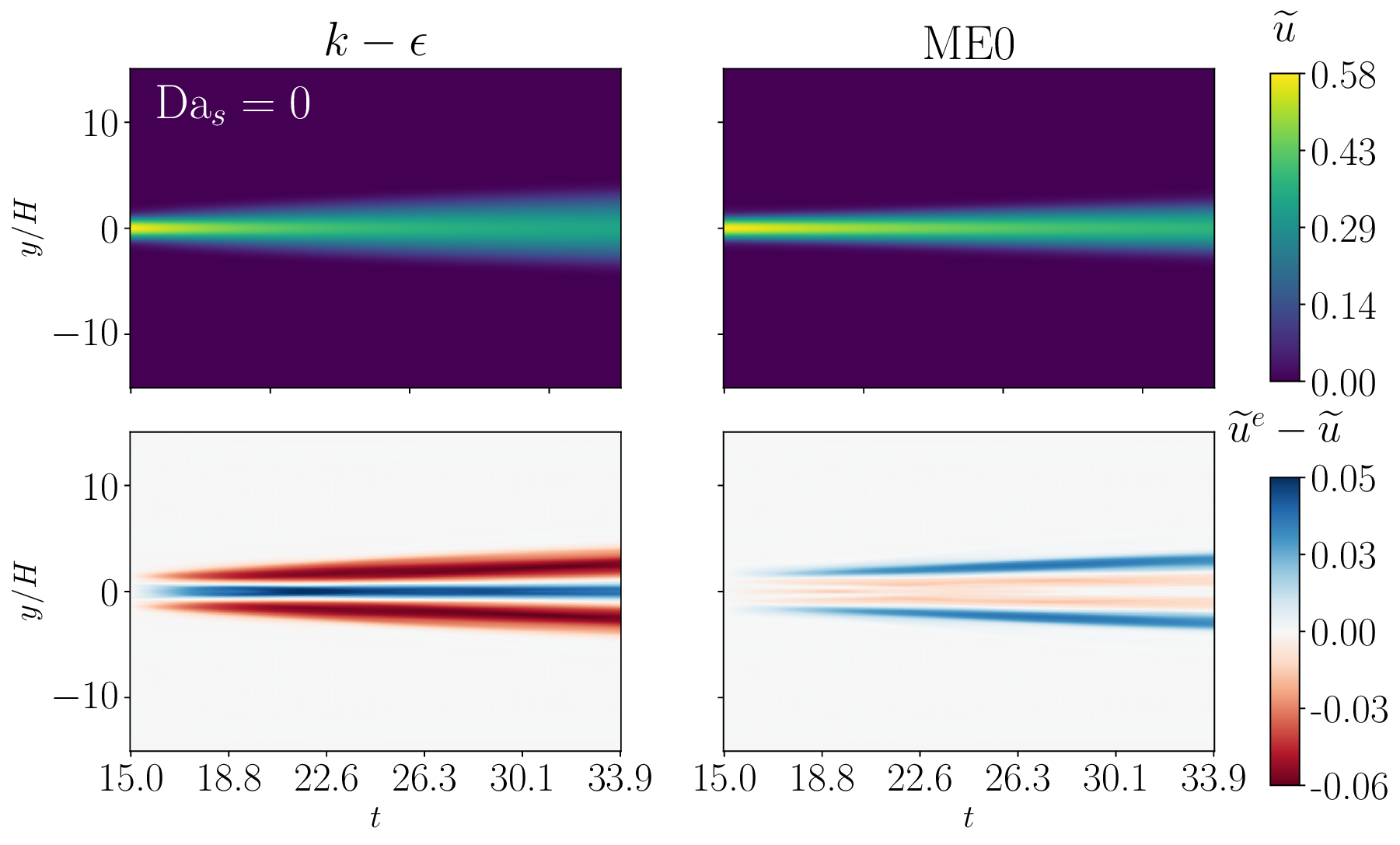}
  \hfill
  \includegraphics[width=0.49\linewidth]{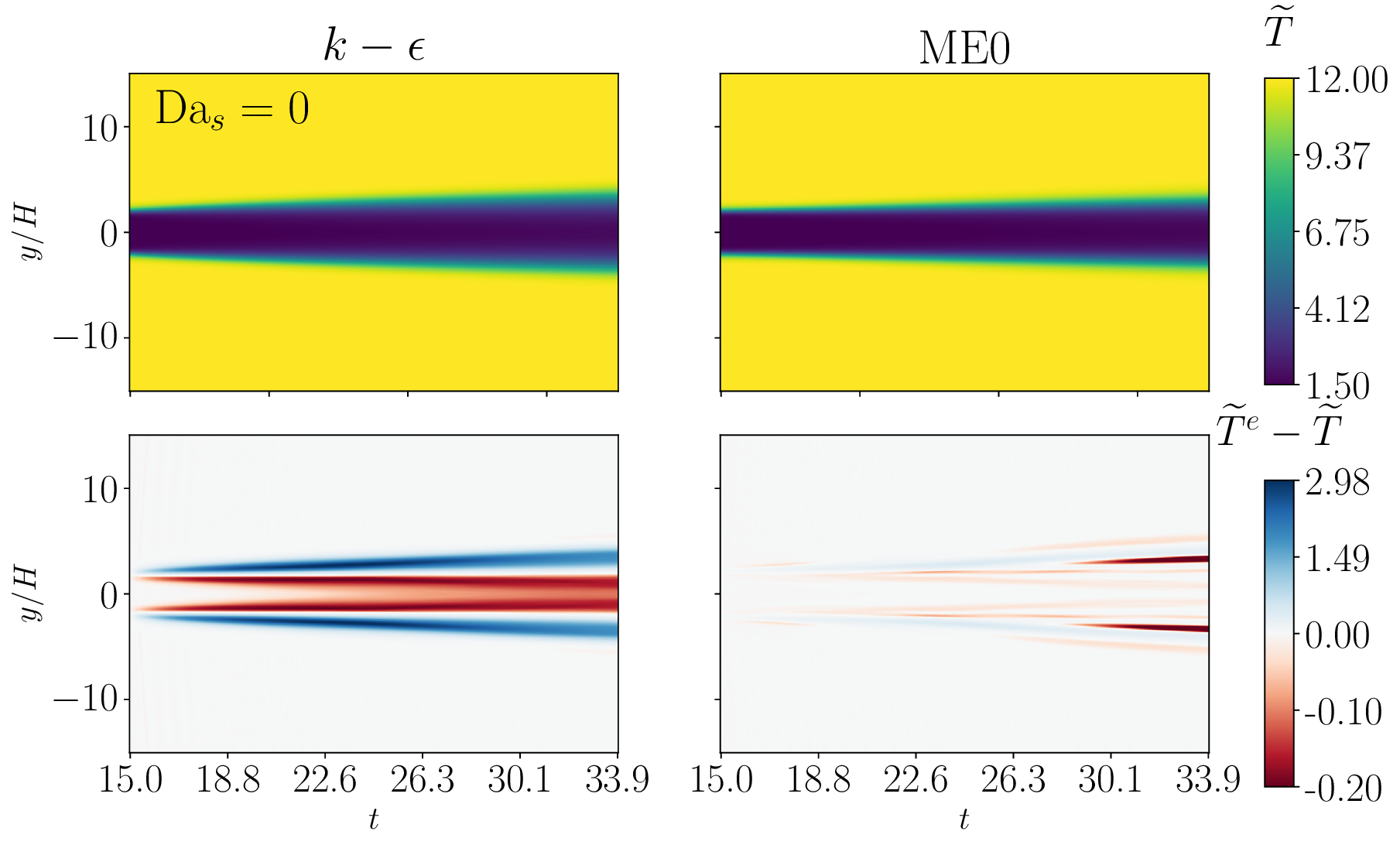}
  \caption{Clockwise from upper left: Cross-stream velocity, density, temperature, and streamwise velocity predictions (top subfigures) and errors (bottom subfigures) for \keps\ RANS (left subfigures) and in-sample DL-augmented RANS (right subfigures). All are shown for $\Das=0$.}
  \label{fig:u-and-T-Da0} 
\end{figure}

In-sample predictions assess model performance for training and testing at the same Damk\"ohler number.
Figure~\ref{fig:u-and-T-Da0} compares RANS predictions of the velocity, density, and temperature using the $k$--$\epsilon$ and ME0-trained models for the $\Da_{s} = 0$ jet. The figures plot the time evolution along the abscissa and the flame-normal mesh along the ordinate. Errors shown in the bottom rows of the figures are the signed differences between the target DNS statistics and the RANS predictions.

\begin{figure}
    \centering
    \includegraphics[width=0.49\linewidth]{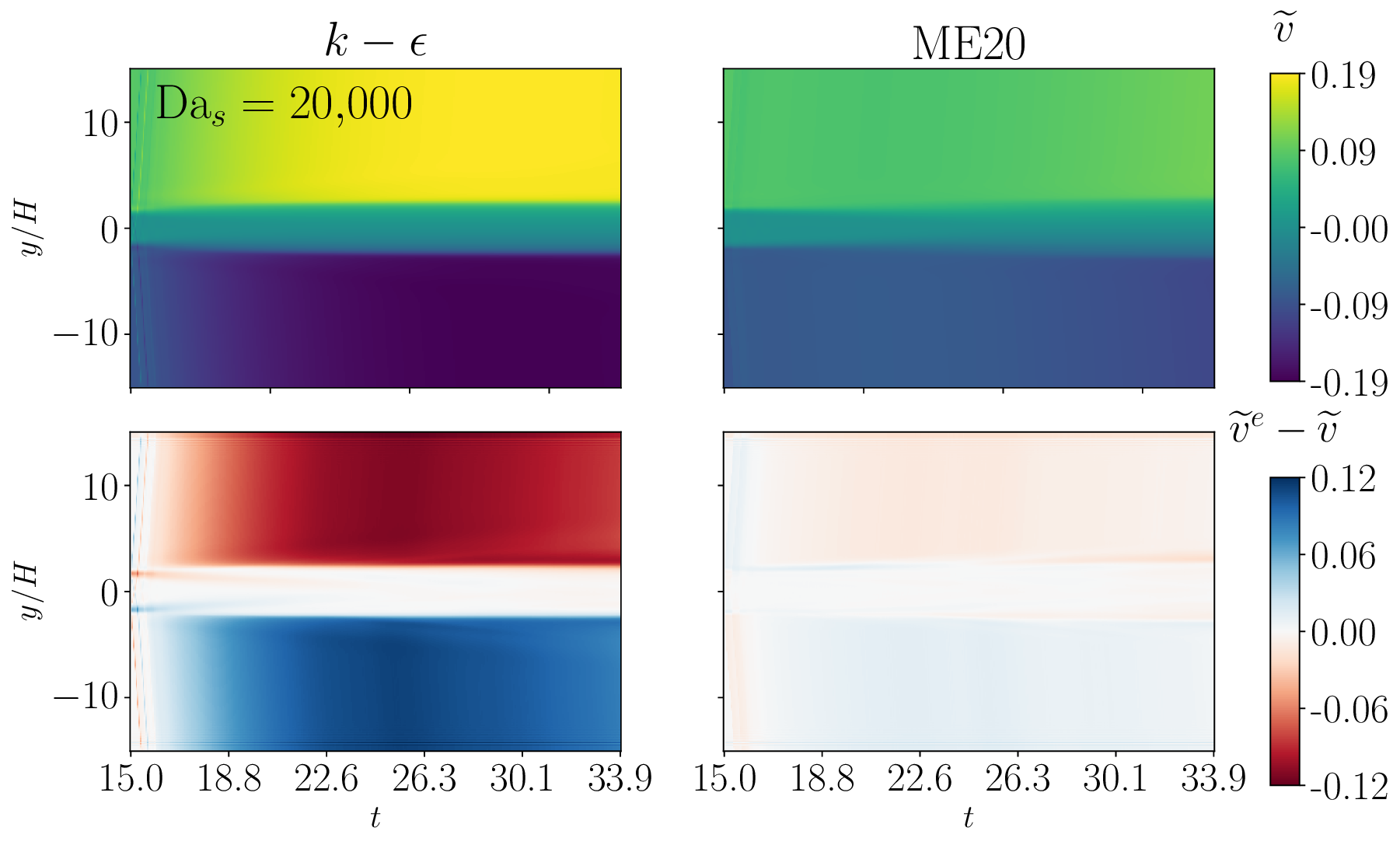}
    \hfill
    \includegraphics[width=0.49\linewidth]{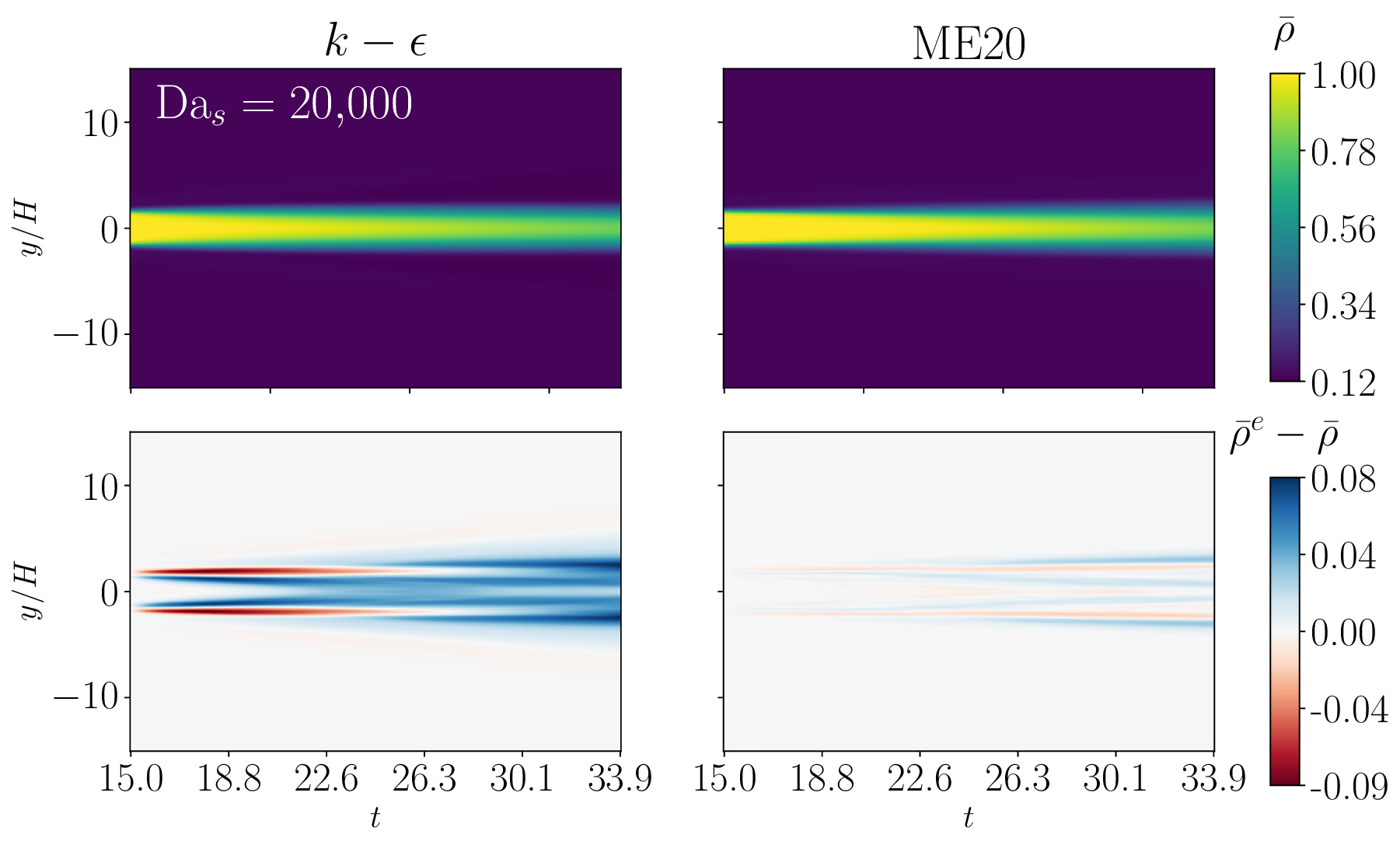}
    \caption{Turbulent flame speed (left) and density (right) predictions and errors for $\Da_{s}= 20{,}000$. }
    \label{fig:v-and-p-Da20k}
\end{figure}

In Figure~\ref{fig:u-and-T-Da0},
the errors for ME0-trained models are magnified ten times, highlighting the superior performance of the DL model over the full time history. Small errors in the ME0-augmented RANS are present at the jet shear layer, though errors in the jet core are negligible.  The streamwise velocity prediction using the $k$--$\epsilon$ model is too diffuse due to its overly large turbulent viscosity, while the errors with the ME0 model are much improved. At later times, the temperature prediction becomes slightly worse for the ME0-augmented RANS model, but it still improves over the baseline $k$--$\epsilon$ model. Overall, the ME0-augmented RANS has significantly lower global errors for all variables compared to the baseline $k$--$\epsilon$ RANS.

To demonstrate the predictive capabilities of the DL-augmented RANS models for turbulent combustion, we train the ``ME" model for $\Da_s = 20{,}000$. Figure~\ref{fig:v-and-p-Da20k} illustrates that this ME20 model significantly improves predictions of turbulent flame speed, $\widetilde{v}$, over the baseline \keps\ model. These improvements result from enhanced temporal accuracy in the evolution of the flame-normal momentum equation~\eqref{eq:rans_v}, facilitated by better predictions of primitive variables such as $\widetilde{T}$ and $\bar{\rho}$ through the equations of state~\eqref{eq:dimensionless_caloric_eos} and~\eqref{eq:eqn_state}. These results underscore the model's ability to accurately capture the tightly coupled dynamics of reacting flows. 

\begin{figure}
\centering
\includegraphics[width=0.75\linewidth]{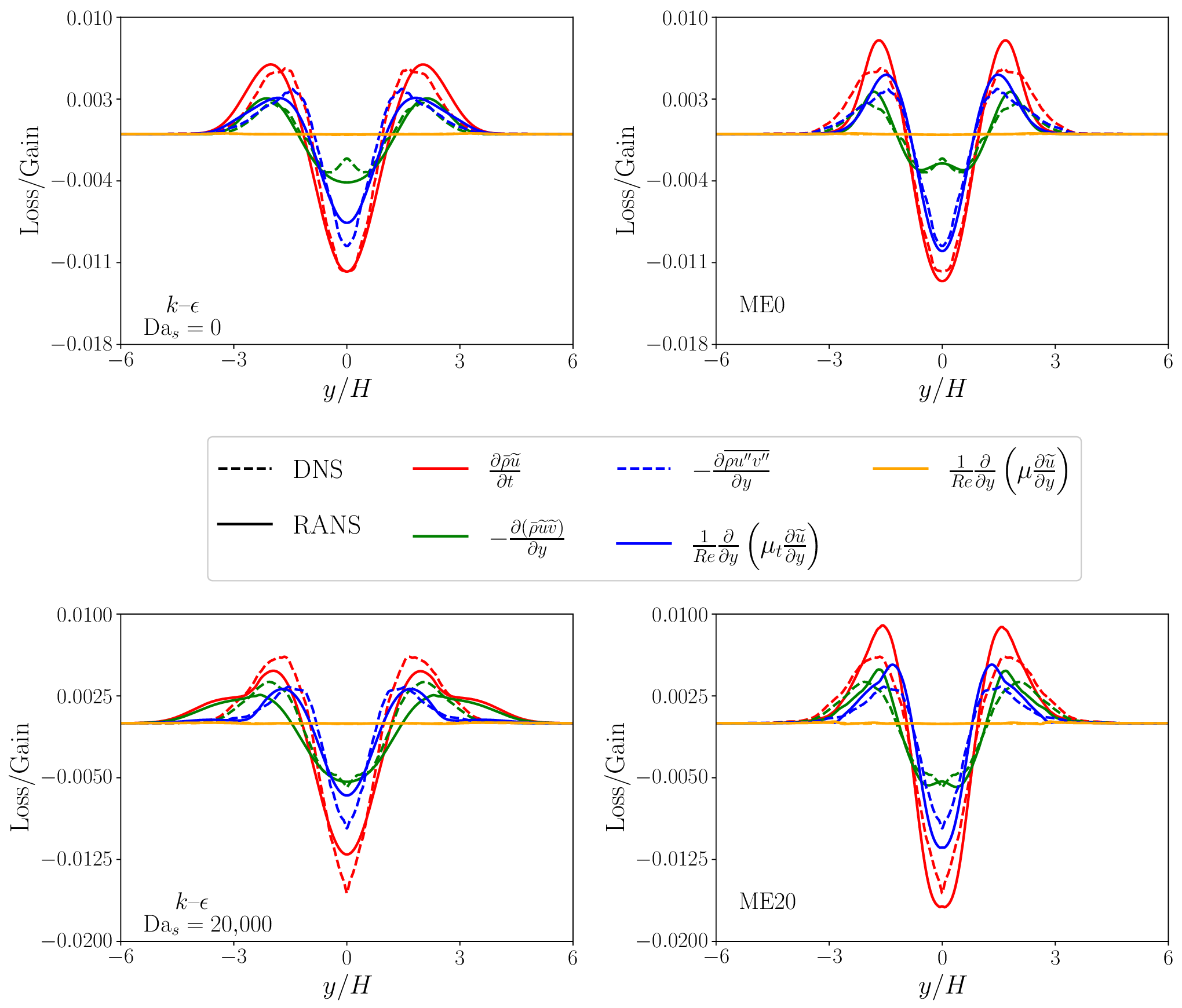}
\caption{Streamwise momentum equation \eqref{eq:rans_u}  budgets for $k$--$\epsilon$ RANS (left)  and DL-augmented RANS (right) predictions for $\Da_{s}= 0$ (top) and $\Da_{s}= 20{,}000 $ (bottom) at $t=25$.} 
\label{fig:closure_mom}
\end{figure}

We next investigate the reasons for the DL-augmented RANS models' improved predictive accuracy. Figure~\ref{fig:closure_mom} plots budgets of the terms in the RANS streamwise momentum equation \eqref{eq:rans_u} from both \emph{a priori} averaged DNS data and \emph{a posteriori} RANS predictions. For $\Da_{s} = 0$ (top row of Figure~\ref{fig:closure_mom}), the DL-augmented Reynolds stress divergence (dashed blue line) more closely approximates the DNS-evaluated Reynolds stress divergence (solid blue line) than does the \keps-modeled Reynolds stress, and the \emph{a posteriori} RANS time derivative more closely matches that of the \emph{a priori} averaged DNS, indicating more closely conforming flow evolution.
At $\Da_{s} = 20{,}000$ (bottom row of Figure~\ref{fig:closure_mom}), the ME20 Reynolds stress divergence is again a closer match to the DNS-evaluated Reynolds stress divergence than the \keps\ closure, though the alignment with the DNS in this reacting case is less clear than in the nonreacting case, in part due to the neglected chemical source term fluctuations. Nonetheless, the ME20-augmented time derivative for $\Da_s=20{,}000$ more closely matches the DNS-evaluated time derivative than does the baseline \keps\ RANS prediction at the jet core.

\begin{figure}[h]
	\centering
	\includegraphics[width=0.75\linewidth]{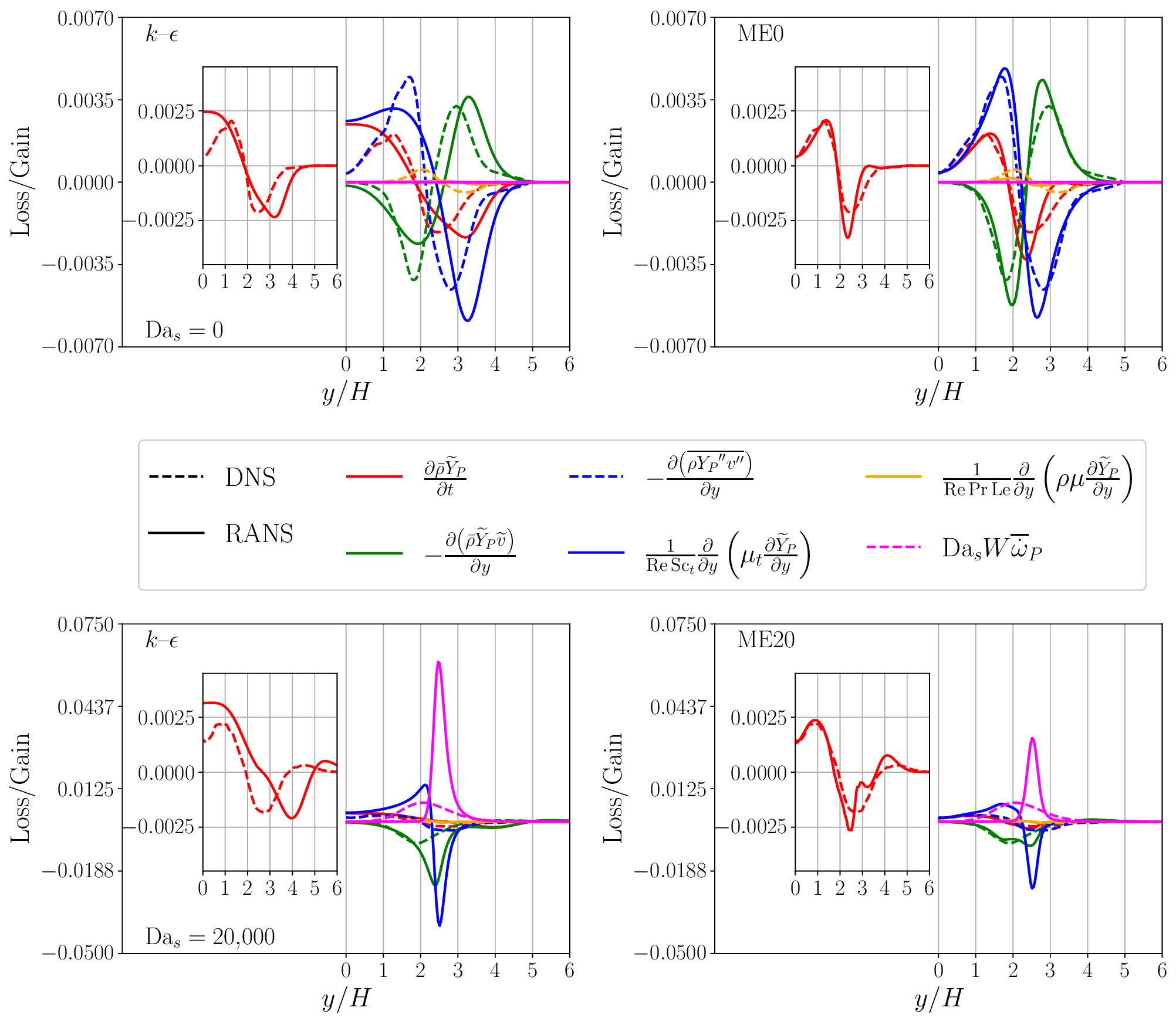}
	\caption{Product species equation (\ref{eq:rans_Y}) budgets for \keps\ RANS (left)  and DL-augmented RANS (right) predictions for $\Da_{s}= 0$ (top) and $\Da_{s}= 20{,}000$ (bottom) at $t=25$. Insets show time derivative terms for clarity in the $\Da_{s}= 20{,}000$ case.
        } 
	\label{fig:closure-species_20k_combined}
\end{figure}

Budgets of the species transport equation~\eqref{eq:rans_Y} are plotted in Figure~\ref{fig:closure-species_20k_combined}. For $k$--$\epsilon$ RANS predictions at $\Da_{s} = 0$ (top row, left), the modeled scalar flux (solid blue line) fails to capture the DNS-evaluated trend (dashed blue line) within the jet core, while the DL-augmented prediction (top row, right) accurately models these closure terms. As a result, the time derivative term (red line) in DL-augmented RANS system more closely agrees with the DNS solution. 

For the reacting case at $\Da_{s} = 20{,}000$ (bottom row of Figure~\ref{fig:closure-species_20k_combined}), errors in the time derivative term for \keps\ RANS are more pronounced than for the nonreacting case, resulting in overall higher errors with the baseline model (see Figure~\ref{fig:v-and-p-Da20k}). The largest sources of error are due to the unclosed scalar flux and the additional error due to the noncommutativity of averaging with the chemical source term (see Section~\ref{subsec:RANS_formulation}). While it does not fully correct these errors individually, the net effect of the ME20 model is to more closely capture the overall time evolution within the jet core than the \keps\ model (see insets in Figure~\ref{fig:closure-species_20k_combined}). Indeed, this is a demonstration of the embedded optimization method's ability to correct the dynamical evolution of the targeted variables (here, the conserved variables $Q$) without necessarily obtaining the DNS closure term, at least without additional constraints on the optimization procedure. Whether this property is desirable or not depends on the modeling objectives at hand.

\begin{figure}
    \centering    \includegraphics[width=0.4\linewidth]{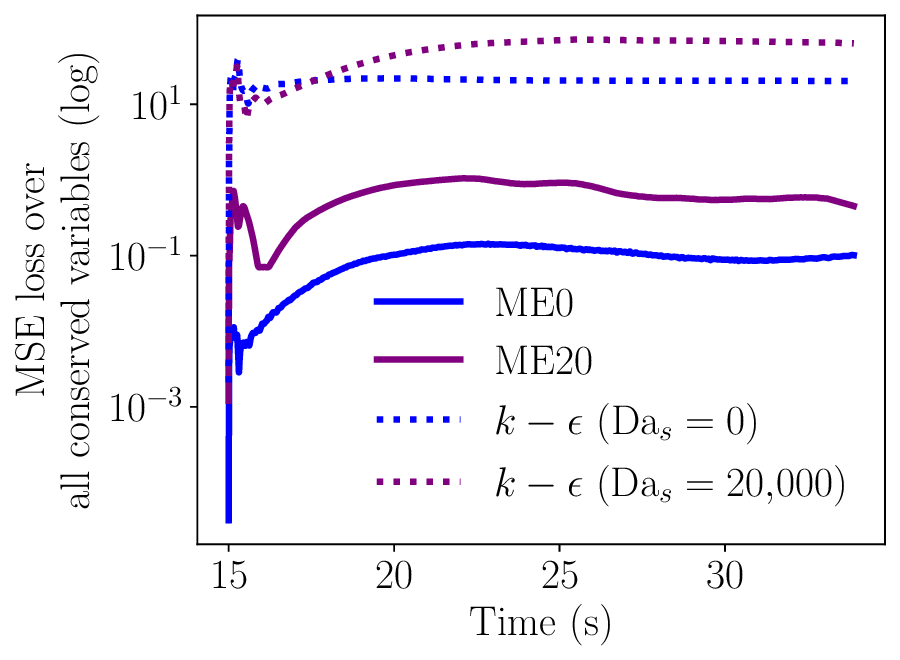}
    \caption{Time evolution of the MSE objective function \eqref{eq:J loss} for \(\Da_s = 0\) and \(\Da_s = 20{,}000\).}
    \label{fig:test_loss}
\end{figure}

Furthermore, the temporal variation of the mean-squared error (MSE) objective function \eqref{eq:J loss} of the DL-augmented RANS predictions reaches a steady state that is approximately two orders of magnitude smaller than that of the corresponding $k$--$\epsilon$ RANS simulations (see Figure~\ref{fig:test_loss}).  This indicates that the DL-augmented RANS framework does not preferentially weight any specific time window and therefore captures the system's temporal evolution in an unbiased manner.

\subsection{Generalization to out-of-sample Damk\"ohler numbers}
\label{subsec:generalization}
\begin{figure}
    \centering
    \includegraphics[width=0.9\linewidth]{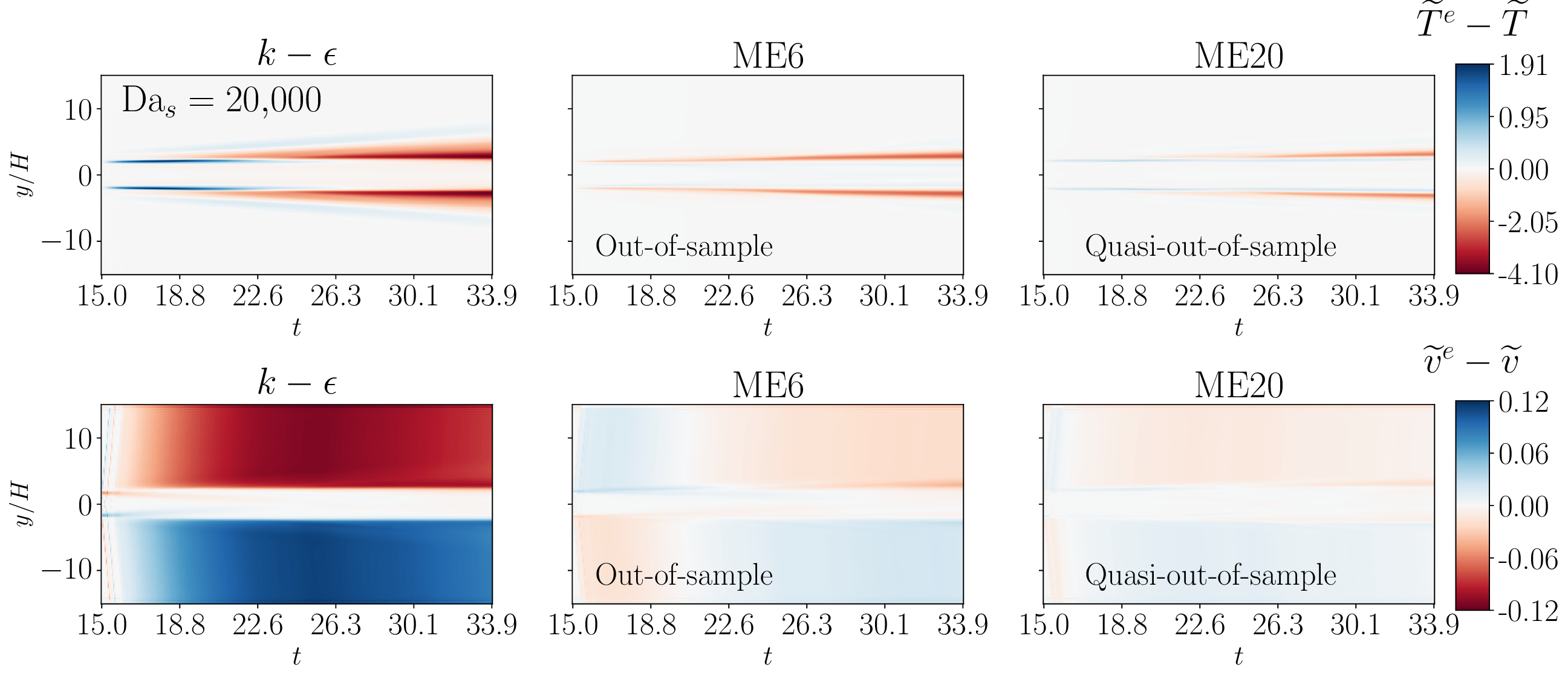}
    \captionsetup{justification=centering}
    \vspace{-0.1cm}
    \caption{Out-of-sample evaluation error in the predicted temperature  (top) and turbulent flame speed (bottom) for $\Da_{s} = 20{,}000$ jet flames using the $k$--$\epsilon$, ME6 (out-of-sample), and ME20 (in-sample) models.}
    \label{fig:Tv_os_Da6k}
\end{figure}

Out-of-sample tests are necessary to evaluate the trained models' stability and accuracy for unseen-in-training scenarios. 
For example, for $\Da_s=20{,}000$ predictions, Figure~\ref{fig:Tv_os_Da6k} compares the $k$--$\epsilon$, ME6 (fully out-of-sample), and ME20 (in-sample) temperature errors. The ME6 prediction qualitatively improves upon that of the $k$--$\epsilon$ model, even though the ME6 training did not involve the $\Da_s=20{,}000$ flow dynamics, while the in-sample case further reduces the errors. 
These lower errors highlight the DL-augmented RANS models' ability to inherently capture the impact of flame dynamics on the mean turbulent flow, as the wrinkling and flame surface area at $\Da_s=20{,}000$ are significantly greater than those at $\Da_s=6{,}000$ (see Figure~\ref{fig:diff_DA}).

\begin{figure}
    \centering
\includegraphics[width=1.0\linewidth]{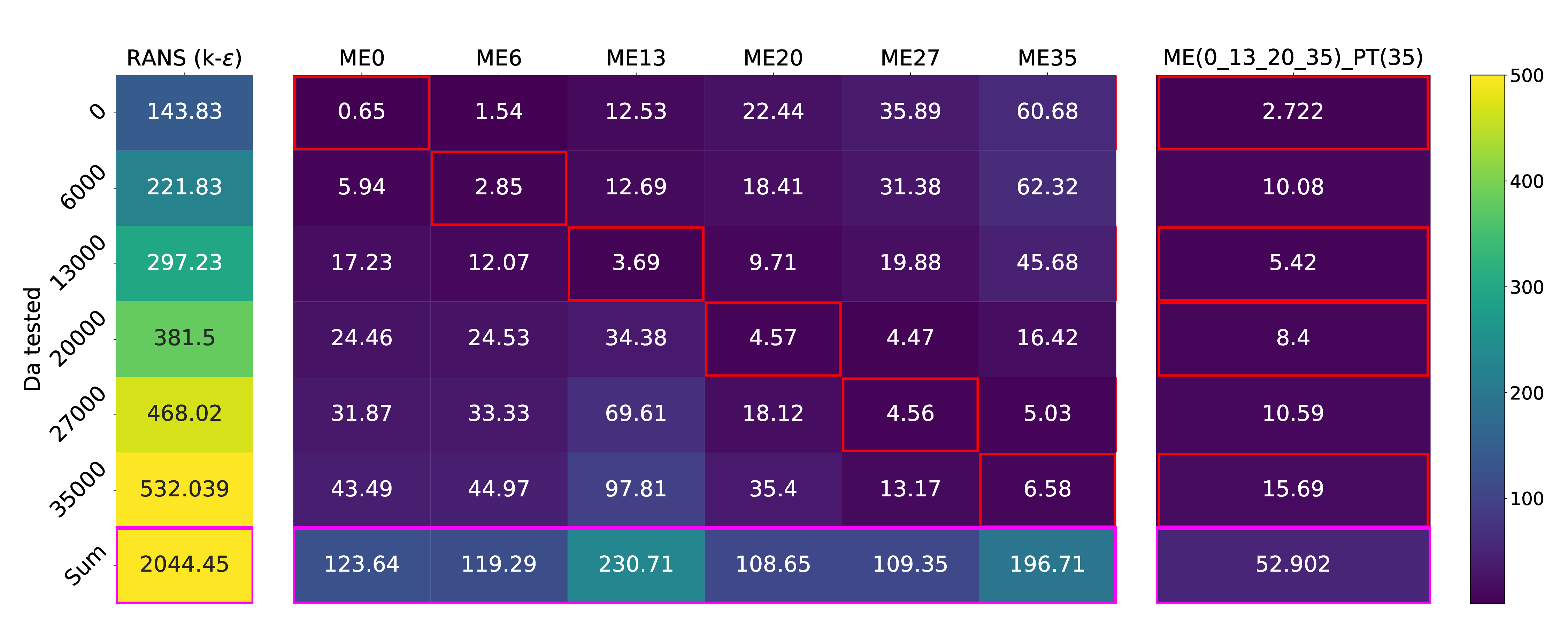}
    \vspace{-2mm}
    \caption{\emph{A posteriori} cumulative (spatiotemporal) MSE loss for different models and testing Damk\"ohler numbers ($\times 1000)$. Computed for $t\in[15,33.9]$. Red boxes indicate in-sample evaluation.}
    \label{fig:map_Da}
\end{figure}

We next assess the out-of-sample performance of DL-augmented RANS by testing each trained model for each scaling Damk\"ohler number. Figure~\ref{fig:map_Da} shows the train--test matrix with cumulative MSE loss over the $N_T$ testing steps, with diagonal entries comprising in-sample evaluation and the remainder representing out-of-sample evaluation.  The first column shows the MSE loss for the standard $k$--$\epsilon$ model, and the remaining the columns report the DL-augmented RANS losses. 

We first assess the out-of-sample performance of single-Damk\"ohler-number-trained models. The ME0 (nonreacting)-trained RANS model significantly outperforms the  baseline $k$--$\epsilon$ closure across the full range of testing Damk\"ohler numbers, even the strongly burning cases. Similarly,  the ME35 model has significantly lower error than the baseline \keps\ model (up to two orders of magnitude lower near its training condition). Evaluated for the nonreacting $\Das=0$ case, the ME35 model is 57.8\,\% more accurate than the \keps\ model. The out-of-sample, \emph{a posteriori} errors of the DL-augmented models significantly improve upon the baseline $k$--$\epsilon$ predictions. Furthermore, the single-Damk\"ohler number-trained models generally exhibit monotonically increasing error for testing Damk\"ohler numbers departing from their in-sample training. Finally, among the single-Damk\"ohler number-trained models, the summed error (last row of Figure~\ref{fig:map_Da}) is lowest for the $\Da_s = 20{,}000$-trained model, which is trained nearly midway along the Damk\"ohler number range. 

To assess generalizability and data scalability of multiple-condition training, the model trained in parallel for $\Da_s \in \{0, 13{,}000, 20{,}000, 35{,}000\}$ is also evaluated in Figure~\ref{fig:map_Da}. While this model has overall higher error for any individual Damk\"ohler number than the corresponding in-sample (single-$\Da_s$-trained model), its net error across the entire testing Damk\"ohler number range is lower than the other models. This illustrates the strong generalizability of models trained across a representative sample of testing configurations. As  the training data size is increased for this multiple-condition-trained model, the overall summed loss decreases from that of the single-condition-trained models, exhibiting data scalability.

Despite their success, all trained DL models will be challenged when applied to out-of-sample conditions. This is most apparent for the $\Das=35{,}000$-trained model tested for $\Das=0$, where its testing error (60.68) is 93 times that of the in-sample, $\Das=0$-trained model (0.65). (We note that the $\Das=35{,}000$-trained model still achieves a 2.5-fold error reduction for the $\Das=0$ test compared to the baseline $k$--$\epsilon$ model.) This can be alleviated across a fixed parameter range by training using data for additional conditions, as we have done for multiple-$\Das$-trained models, though the challenge of generalizing to unseen conditions remains.

\subsection{Influence of weight initialization on test errors}
\label{subsec:weight_init}
\begin{figure}
    \centering
\includegraphics[width=1\linewidth]{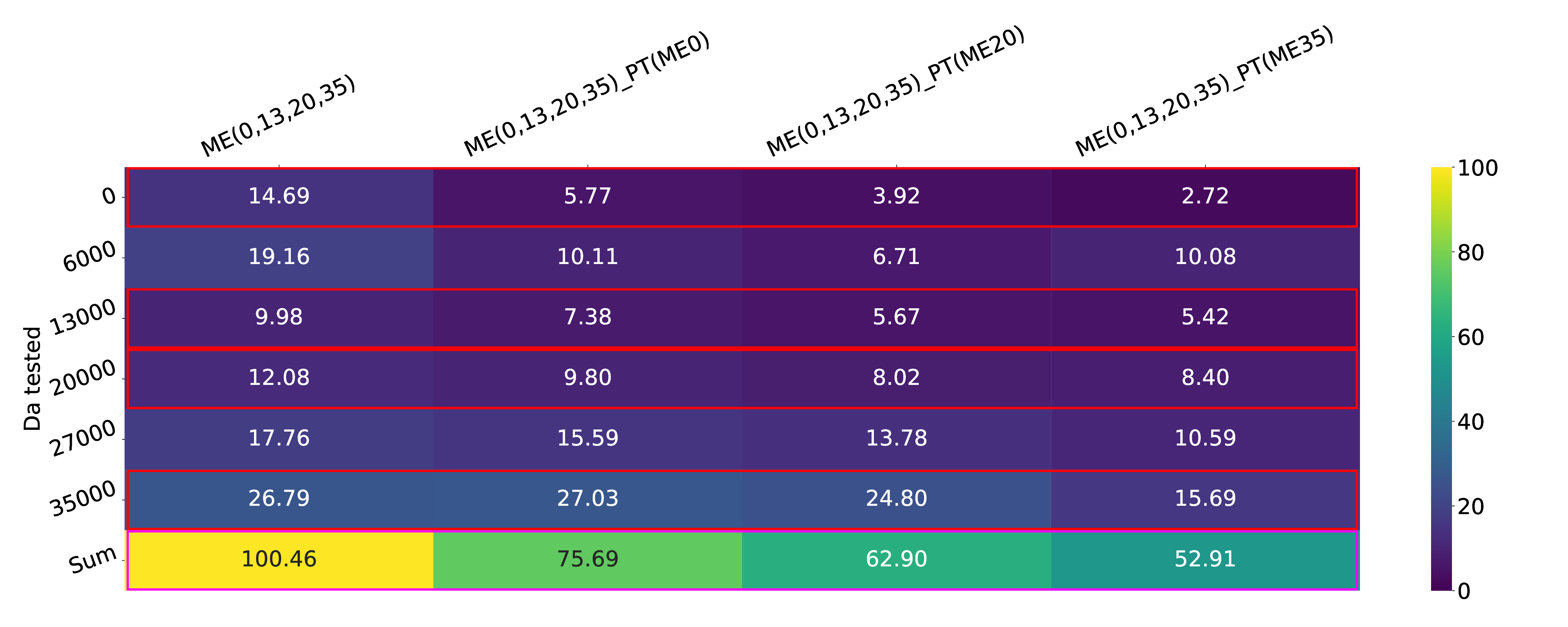}
    \vspace{-2mm}
    \caption{\emph{A posteriori} cumulative (spatiotemporal) MSE loss for multiple-Damk\"ohler number-trained models with different weight initializations. Computed for $t \in [15, 33.9]$. Red boxes indicate in-sample evaluation.}
    \label{fig:map_Da_MPI}
\end{figure}

The effects of weight initialization for multiple-Damk\"ohler number-trained models are evaluated in Figure~\ref{fig:map_Da_MPI}. It shows the cumulative MSE loss of models trained with random weight initialization (first column), ME0 parameters (second column), ME20 (third column) and ME35 parameters (fourth column). The randomly initialized model has the highest overall loss, while the model initialized from ME0 has the highest test loss for $\Da_s = 35{,}000$, suggesting that higher $\Da_s$ values do not train as effectively with ME0 pretraining (see also Figure~\ref{fig:train_loss_combined}) (center). Finally, the model initialized from the ME35 parameters yields the lowest out-of-sample and overall summed errors. These results emphasize the importance of inductive bias in pretrained models. Consequently, initializing from a model trained on a more challenging or representative condition, such as the $\Da_s=35{,}000$ case among these jet flames, is advantageous for parallel training across a wide range of conditions.

\section{Discussion}
\label{sec:discussion}

\subsection{Computational cost of adjoint-based training and inference}

We now evaluate the computational cost of the adjoint-based training approach. For the present 1D, unsteady RANS predictions, the baseline \keps\ calculations,  DL-augmented RANS model training, and DL-augmented RANS inference were performed on an AMD Ryzen Threadripper PRO 5945WX (4.1 GHz) using 12 CPU threads. Table~\ref{tab:comp_cost} lists typical walltimes per epoch for these three types of calculations, the number of epochs used per calculation (one for inference; 200 for model training), and total walltime in hours. The DL-augmented RANS inference requires 2.2 times the cost of a \keps\ RANS prediction, largely due to the cost of evaluating the neural network, while the cost for one epoch of DL-augmented RANS training is approximately 5 times that of inference due to the adjoint pass and largely comprises the cost of evaluating Jacobians.

While not negligible, the training cost is not intractable: the adjoint cost per epoch is of the same order of magnitude as the forward prediction cost, and training does not require significantly greater CPU or memory resources than the prediction; therefore, the same hardware (e.g., number of CPU cores or number of GPUs) can be used for DL-augmented RANS training as for \keps\ RANS predictions. The main cost driver for training is the number of epochs required to converge the model. While the walltime training cost (for 200 epochs) is significantly greater than that of a single prediction, training a model is a one-time cost that can be amortized over many predictions.

\begin{table}
    \centering
    \caption{Computational cost of RANS inference and testing}
    \begin{tabular}{lccc}
        \toprule
         & {Walltime per epoch (s)} & {Epochs} & {Total walltime (hours)} \\
        \midrule
        \keps\ RANS & 68.96 & 1 & 0.0191 \\
        DL-RANS inference & 150.92 & 1 & 0.0419 \\
        DL-RANS training & 768.18 & 200 & 42.67 \\
        \bottomrule
    \end{tabular}
    \label{tab:comp_cost}
\end{table}

\begin{figure}
    \centering    
    \includegraphics[width=0.55\linewidth]{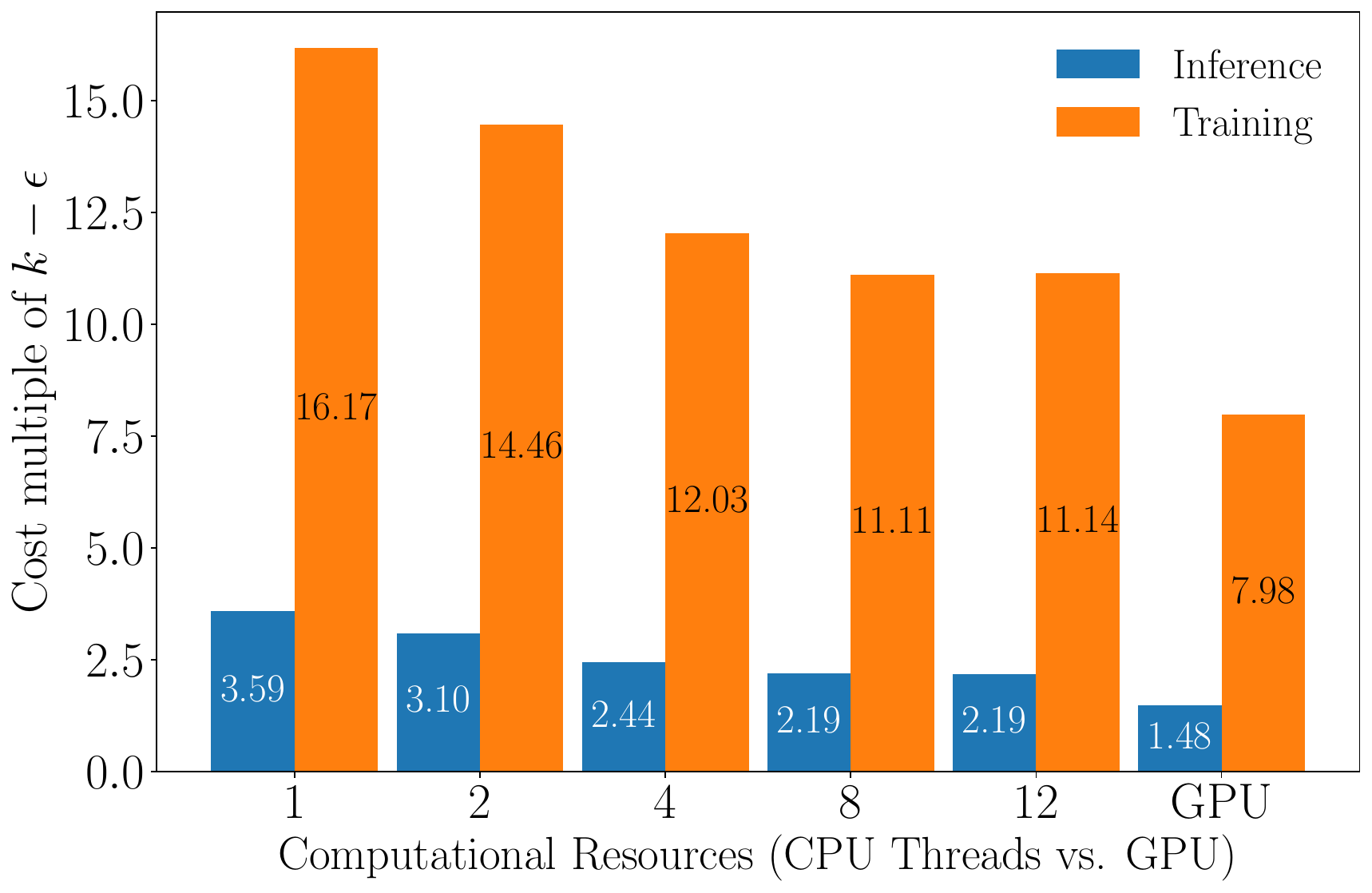}
    \caption{Computational cost of DL-augmented RANS training and inference, for increasing CPU threads and GPU acceleration, relative to the baseline \keps\ RANS prediction.}
    \label{fig:comp_cost}
\end{figure}

Figure~\ref{fig:comp_cost} illustrates the effect of CPU multithreading and GPU acceleration (using an NVIDIA RTX A2000 GPU) on the performance of the DL-augmented model training and inference compared to the baseline \keps\ RANS evaluation on the same resources. Even for the present 1D-unsteady RANS problem, increasing the CPU threads modestly improves the efficiency of the DL model training calculations, up to 8 threads, compared to the baseline \keps\ RANS due to the vectorizability of neural network backpropagation. GPU evaluation further increases training and inference efficiency over the values reported in Table~\ref{tab:comp_cost} due to the GPU's increased vector register width. For 2D RANS, and especially 3D LES, the efficiency of neural network operations on GPUs means that the additional cost of evaluating the ML model will be marginal compared to to the higher baseline prediction cost.

\subsection{Comparison of DPM to \emph{a priori} models}
\label{sec: alternate_optimization}
We assess the trade-offs between PDE-constrained embedded optimization (DPM) and standard \emph{a priori} (offline) modeling in two steps. As a ``best-case'' scenario, we first evaluate the performance of RANS using $\mu_t$ obtained directly from the DNS-evaluated $k$ and $\epsilon$. Such an approach would clearly be infeasible for general-purpose RANS, as it requires DNS data for the configuration of interest.
However, it permits evaluation of model-consistency errors (due to the fact that the \emph{a priori} model is not optimized consistently with the RANS equations) independently of any model-form errors (for example, inadequacy of a neural network fit to data). Second, we optimize a neural network model to the \emph{a priori}, DNS-evaluated $\mu_t$ as an example of a generalizable \emph{a priori} modeling approach.

\begin{figure}
    \centering
    \includegraphics[width=0.45\linewidth]{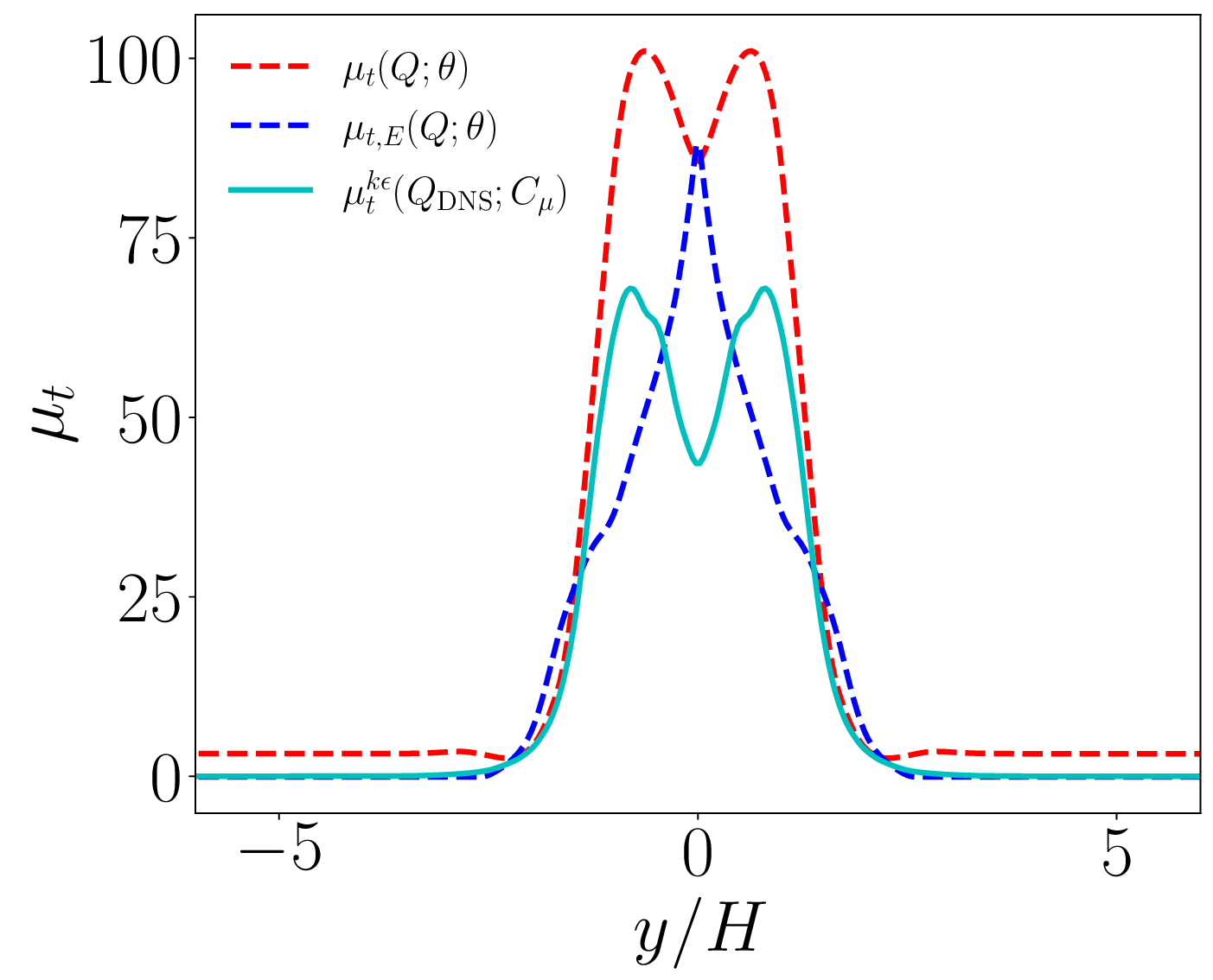}
    \vspace{-0.1cm}
    \caption{Comparison of DPM-optimized eddy viscosity and eddy diffusivity using the ME0 model to  \emph{a priori}, DNS-evaluated $k$--$\epsilon$ eddy viscosity. Shown for $\Da_s=0$ at $t = 15$.}
    \label{fig:DNS_mut}
\end{figure}

We first evaluate the turbulent viscosity directly using the \emph{a priori}, DNS-evaluated TKE and dissipation rate using \eqref{eq:rans_mut_kep}; we designate this closure $\mu_t^{k\epsilon}(Q_\DNSsub;C_{\mu})$ with $C_\mu=0.09$ as before. The resulting eddy viscosity is comparable in magnitude to the DPM-optimized, ME0-modeled $\mu_t(Q;\theta)$, shown in Figure~\ref{fig:DNS_mut} for $\Da_{s}=0$. The \emph{a priori}, DNS-evaluated turbulent thermal diffusivity $\mu_{t,E}^{k\epsilon}(Q_\DNSsub;C_\mu)$ is computed using $\Pr_t=0.9$.

We additionally train a neural network for the \emph{a priori} eddy viscosity using a formulation analogous to the DPM model \eqref{eq:NN formulation},
\begin{equation}
  \label{eq:NN_ap_formulation}
  \mu_{t,ap}(\theta) \leftarrow \mathcal{N}_{ap}\left( \frac{\partial \bar{\rho}}{\partial y},\, \frac{\partial \bar{\rho} \widetilde{u}}{\partial y},\, \frac{\partial \bar{\rho}\widetilde{v}}{\partial y},\, \frac{\partial \bar{\rho}\widetilde{E}}{\partial y},\, \frac{\partial \bar{\rho} \widetilde{Y}_P}{\partial y},\, \frac{\partial \bar{p}}{\partial y},\, \Da_s\, \dot{\omega}_{P}(\overline{\rho},\widetilde{Y_{P}},\widetilde{T}),\, t \right),
\end{equation}
with the turbulent thermal diffusivity likewise obtained using $\Pr_t=0.9$. The architecture of $\mathcal{N}_{ap}$ is the same as that of the DPM networks. We train $\mathcal{N}_{ap}$ over 7,000 DNS time steps across 2,000 epochs, which requires approximately 3 hours on a single NVIDIA RTX A2000 GPU. The training employs the MSE loss
\begin{equation}
  \label{eq:NN_ap_loss}
  J(\theta) = \frac{1}{ N'_{y}}  \sum_{i=1}^{N'_{y}} \left(\left(\mu_{t,ap}(Q_\DNSsub;\theta) - \mu_{t}^{k\epsilon}(Q_\DNSsub;C_{\mu})\right)^2\right)(y_{i}).
\end{equation}
The resulting $\mu_{t,ap}(Q;\theta)$ closes the RANS equations. As with any \emph{a priori} model, inference in \emph{a posteriori} simulations uses the predicted $Q$ rather than the DNS-evaluated $Q_\DNSsub$ used for training, which is a potential source of model-consistency error.

\begin{figure}
    \centering
    \includegraphics[width=\linewidth]{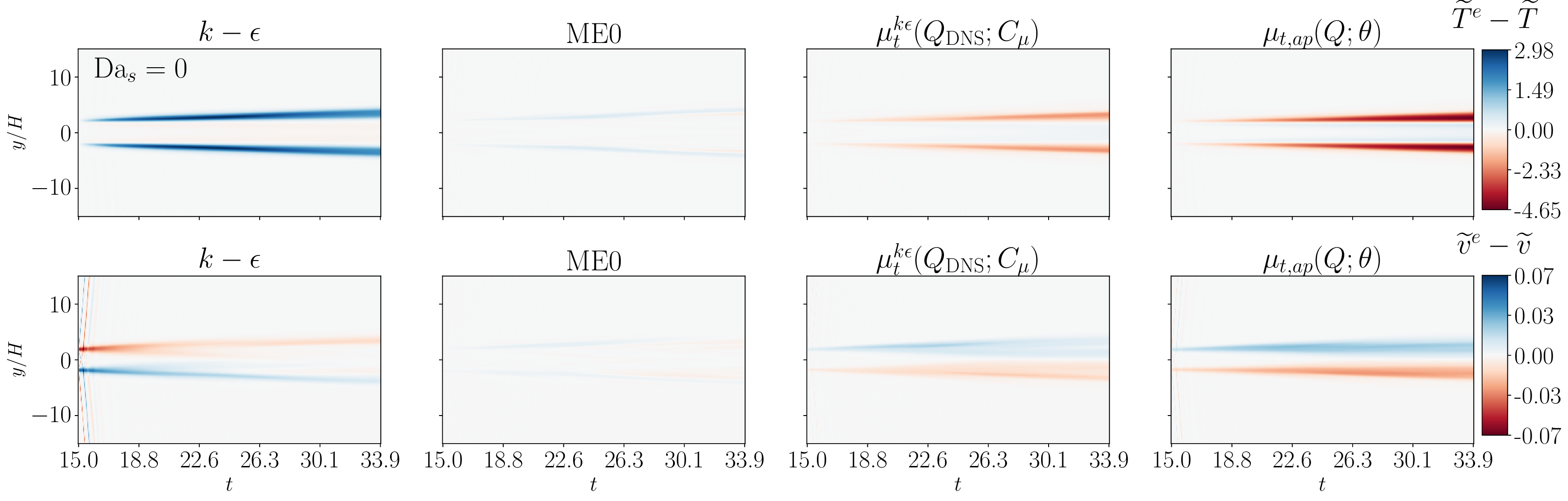}
    \vspace{-0.1cm}
    \caption{\emph{A posteriori} RANS errors in predicted temperature (top) and cross-stream velocity (bottom) for $\Da_{s} = 0$ jets using the $k$--$\epsilon$ model, DPM-optimized ME0 model, direct DNS-evaluated $\mu_{t}^{k\epsilon}(Q_\DNSsub;C_{\mu})$, and \emph{a priori}-modeled $\mu_{t,ap}(Q;\theta)$.
    }
    \label{fig:Tv_Da0_DNS}
\end{figure}

Figure~\ref{fig:Tv_Da0_DNS} compares the errors in the \emph{a posteriori} RANS temperature and cross-stream velocity, with respect to the DNS target data, using the $k$--$\epsilon$ model, DPM-optimized ME0 model, direct DNS-evaluated $\mu_{t}^{k\epsilon}(Q_\DNSsub;C_{\mu})$, and \emph{a priori}, neural network-modeled $\mu_{t,ap}(Q;\theta)$ for the baseline \(\Da_{s}=0\) case.
For both \emph{a priori} closures, the in-sample errors are significantly higher than those of the DPM-optimized ME0 model.

Table~\ref{tab:test_error_mut_DNS} lists the cumulative MSE errors for these four models for the in-sample $\Da_s=0$ case and out-of-sample predictions at $\Da_s=20{,}000$. In general, RANS predictions using the direct DNS-evaluated $\mu_{t}^{k\epsilon}$ are more accurate than the corresponding \keps\ RANS, as could be anticipated for the in-sample $\Da_s=0$ case, though this also holds when the $\Da_s=0$ DNS-evaluated eddy viscosity is used for out-of-sample predictions at $\Da_{s}=20{,}000$. The \emph{a priori}-optimized neural network model is comparably accurate to the baseline \keps\ model in-sample but is unstable when evaluated for $\Da_{s}=20{,}000$. In both cases, the DPM-optimized ME0 model is more than one order of magnitude more accurate than each of the comparison models.

\begin{table}
    \centering
    \begin{tabular}{ccccc}
        \toprule
        & \multicolumn{2}{c}{$\Da_{s}=0$} & \multicolumn{2}{c}{$\Da_{s}=20{,}000$} \\
        Model  & MSE & Rel.~Error (\%) & MSE & Rel.~Error (\%) \\
        \midrule
        $k$--$\epsilon$ & 143.8 & 100.0 & 381.5 & 100.0 \\
        ME0 & 0.65 & 0.45 & 22.44 & 5.88 \\
        $\mu_{t}^{k\epsilon}(Q_\DNSsub;C_{\mu})$ & 39.81 & 27.7 & 145.43 & 38.1 \\
        $\mu_{t,ap}(Q;\theta)$ & 150.85 & 104.9 & -- & -- \\
        \bottomrule
    \end{tabular}
    \caption{Cumulative MSE errors of RANS using the \keps\ model, the DPM-optimized ME0 model, direct DNS-evaluated $\mu_t^{k\epsilon}$, and the \emph{a priori}-trained $\mu_{t,ap}$ model tested \emph{a posteriori} for $\Da_{s}=0$ and $\Da_{s}=20{,}000$. The direct DNS-evaluated $\mu_t^{k\epsilon}$ is taken from $\Da_s=0$ for both testing Damk\"ohler numbers. Relative errors are shown with respect to the baseline $k$--$\epsilon$ model. The $\Da_{s}=0$-trained $\mu_{t,ap}(Q;\theta)$ model is unstable for $\Da_{s}=20{,}000$.}
    \label{tab:test_error_mut_DNS}
\end{table}

The superior performance of the DPM-optimized model over \emph{a priori} methods can be attributed to the former's inherent \emph{model consistency}~\cite{duraisamy2021perspectives}, which DPM achieves by accounting for feedback between the model and the governing PDEs during training. \emph{A priori} optimization does not account for this feedback loop, which leads to poor predictive performance and potential instability in \emph{a posteriori} applications~\cite{sirignano2020dpm,macart2021embedded}. In the present study, the DPM approach is uniquely able to account for the influence of the RANS $k$ and $\epsilon$ variables on predictions, which is a significant advantage, as these variables are defined differently in RANS (in which they are the solution of PDEs) than they are in DNS (in which they are statistics of the instantaneous flow field). This is a feature of DPM that distinguishes it from the state of the art.

Another advantage of DPM is its ability to optimize over objective functions obtained from any PDE solution variables, including derived quantities such as spatially filtered and/or time-averaged statistics \cite{macart2021embedded,liu2024adjoint}. This enables the DPM optimization procedure to target experimental data, from which the closure terms needed by \emph{a priori} optimization may not be readily available, or even data from ``higher-order'' simulations such as direct simulation Monte Carlo solutions of the Boltzmann equation \cite{nair2023deep}, which might lack obvious connections to the modeled closure terms. 
In contrast, \emph{a priori} optimization must directly target the outputs of the closure model and hence is typically limited to DNS target data. Although DPM training is an order of magnitude more expensive than \emph{a priori} methods, its advantages outweigh the training cost, while the inference cost remains comparable for both approaches.

The main challenge of PDE-constrained optimization (not necessarily limited to DPM) is the cost of solving the optimization problem by iterating over the forward PDEs. DPM uses adjoint equations to obtain the gradients needed for optimization, which significantly alleviates the compute and memory requirements of non-graph-breaking alternatives (e.g., \cite{akhare2023physics}). The adjoint equations can be complicated, especially for high-dimensional problems and detailed chemical kinetics. However, modern algorithmic differentiation frameworks, such as that provided by \emph{PyTorch}~\cite{paszke2019pytorch}, have greatly simplified the process of constructing the discrete-exact adjoint PDEs, which has made DPM accessible for large-scale flow simulations.

\section{Conclusion}
\label{sec:conclusion}
Turbulent flow interactions with chemical heat release can invalidate conventional eddy-viscosity closures for RANS and LES, and the severity of these effects varies with the Damk\"ohler number. To address these limitations, and to capture the challenging regime dependence, we develop a deep learning PDE augmentation method (DPM)-based closure for turbulent combustion RANS. The closure leverages the existing compressible $k$--$\epsilon$ formulation while augmenting its capability for reacting flows. Model optimization is directly constrained by the RANS PDEs, and a set of adjoint PDEs are introduced and solved to perform this optimization in a memory-efficient manner.
Training data is obtained from turbulent premixed jet flame DNS at six Damk\"ohler numbers. We simulate the dimensionless, compressible Navier--Stokes equations with single-step kinetics, which reduces computational cost and enables the dimensionless form while adequately introducing heat-release effects on turbulence. These are qualitatively similar to those of detailed-mechanism DNS at high and low Damk\"ohler numbers \cite{macart2018effects}. We further validate the nonreacting DNS against previous experimental and numerical results.

In-sample, \emph{a posteriori} RANS calculations for turbulent premixed planar jet flames demonstrate the significantly improved predictive accuracy of the neural network-augmented equations compared to the baseline $k$--$\epsilon$ equations, with cumulative spatiotemporal errors reduced by several orders of magnitude. This enhanced accuracy can be attributed to two key factors. First, the DL-augmented closures significantly improve the Boussinesq closures for Reynolds stress and scalar fluxes for all testing Damk\"ohler numbers. Second, the DL models are specifically trained to augment the dynamics of the evolution equations. For the nonreacting case, the  scalar flux closure is notably improved over the baseline gradient diffusion closure. In reacting cases, the model additionally compensates for unclosed chemical source terms, leading to more accurate predictions of the overall jet flame dynamics.

Out-of-sample, \emph{a posteriori} predictions using the DL-augmented models also show substantial improvement over the baseline $k$--$\epsilon$ model. This indicates that the models learn generalized underlying functions, which enable extrapolation about the training data. Further generalization and data scaling are shown by training DL-augmented RANS models over multiple Damk\"ohler numbers with different neural network parameter initializations. This results in further reduced \emph{a posteriori} errors across the range of in- and out-of-sample testing Damk\"ohler numbers.

The trained models demonstrate strong out-of-sample stability and generalizability, underscoring the potential of the DPM-based method to achieve predictive accuracy across a broad range of turbulent combustion conditions. Future work could assess the applicability of transfer learning to improve generalization across out-of-sample Damk\"ohler numbers and the ability to generalize across a range of Reynolds numbers, combustor geometries, and nonpremixed/partially premixed combustion regimes. Current efforts are focused on developing analogous models for LES, where in the modeling challenge is compounded by the influence of the filter-scale Damk\"ohler number.

\section*{Acknowledgments}
This work was supported by the U.S.\ National Science Foundation under Award CBET-22-36904. This work used resources of the Oak Ridge Leadership Computing Facility, which is a DOE Office of Science User Facility supported under Contract DE-AC05-00OR22725 (INCITE award ``Extreme-Scale Data Assimilation for Predictive Flow Simulations''). 

\section*{Declaration of Interest}
The authors declare that they have no known competing financial interests or personal relationships that could have appeared to influence the work reported in this paper.

\appendix
\section{Derivation of the RANS total energy equation}
\label{appen:energy_derivation}

The derivation of the dimensionless RANS total energy equation proceeds as follows. We first obtain a dimensional version of the unclosed RANS total energy equation, propose closures, and finally nondimensionalize to obtain \eqref{eq:rans_E}. 

The dimensional, density-weighted (Favre) averaged total energy transport equation (the dimensional version of \eqref{eq:NS_energy}) is
\begin{align}
    \frac{\partial (\bar{\hat{\rho}} \widetilde{\hat{E}})}{\partial \hat{t}} 
    + \frac{\partial}{\partial \hat{y}} \left( \widetilde{\hat{v}} \bar{\hat{\rho}} \widetilde{\hat{E}} \right) 
    +\frac{\partial (\bar{\hat{p}} \widetilde{\hat{v}})}{\partial \hat{y}} 
    -\frac{\partial}{\partial {\hat{y}}}\left(\frac{\hat{C}_{p}\hat{\mu}}{\Pr} \frac{\partial \widetilde{\hat{T}}}{\partial {\hat{y}}} \right)
    - \frac{\partial}{\partial \hat{y}}\left(\frac{\hat{C}_{p}\hat{\mu}}{\Pr} \frac{\partial \overline{\hat{T}^{\prime \prime}}}{\partial {\hat{y}}}\right)
    + \frac{\partial}{\partial \hat{y}} \left( \hat{R} \overline{\hat{\rho}\hat{v}^{\prime \prime}\hat{T}^{\prime \prime}} + \overline{{\hat{\rho} \hat{v}^{\prime \prime} \hat{E}^{\prime \prime}}}  \right)  \notag
    &\\- \frac{\partial}{\partial \hat{y}} \left( \overline{\hat{\tau}_{ij} \hat{u}_{i}^{\prime \prime}} + \overline{\hat{\tau}_{ij}} \widetilde{\hat{u_{i}}}  \right) 
    - \frac{\partial}{\partial \hat{y}} \left( \sum_{k=1}^{N} \bar{\hat{\rho}}\widetilde{\hat{h}}_k D\frac{\partial \widetilde{\hat{Y}}_{k}}{\partial \hat{y}} + \sum_{k=1}^{N}\overline{\hat{\rho} \hat{h}_k^{\prime \prime}D\frac{\partial \hat{Y}_{k}^{\prime \prime}}{\partial{\hat{y}}}}  \right) = 0, \label{eq:dim_energy}
\end{align}
where $\hat{R} = {\hat{R}_{u}}/{\hat{W}_{k}}$ is the specific gas constant,  $\hat{C}_{p} = \gamma \hat{R}/(\gamma-1)$ is the specific heat at constant pressure,   and a constant Prandtl number has been assumed. We first eliminate the total energy fluctuation in the first unclosed term, $\hat{R} \overline{\hat{\rho}\hat{v}^{\prime \prime}\hat{T}^{\prime \prime}} + \overline{\hat{\rho}\hat{v}^{\prime \prime} \hat{E}^{\prime \prime} }$, by expanding it as
\begin{align}
\hat{R} \overline{\hat{\rho}\hat{v}^{\prime \prime}\hat{T}^{\prime \prime}} + \overline{\hat{\rho}\hat{v}^{\prime \prime} \hat{E}^{\prime \prime}} &= 
 \hat{R} \overline{\hat{\rho}\hat{v}^{\prime \prime}\hat{T}^{\prime \prime}} +  \overline{\hat{\rho}\hat{v}^{\prime \prime}\hat{e}^{\prime \prime}} + \overline{\hat{\rho}\hat{v}^{\prime \prime}\hat{u}_{k}^{\prime \prime}\widetilde{\hat{u}}_{k}} +  \frac{1}{2}\overline{\hat{{\rho}}\hat{v}^{\prime \prime}(\hat{u}_{k}^{\prime \prime}\hat{u}_{k}^{\prime \prime})} \notag \\
 &= 
 \hat{R} \overline{\hat{\rho}\hat{v}^{\prime \prime}\hat{T}^{\prime \prime}} +  \overline{\hat{\rho}\hat{v}^{\prime \prime}\hat{e}^{\prime \prime}} + \overline{\hat{\rho}\hat{v}^{\prime \prime}\hat{u}_{k}^{\prime \prime}\widetilde{\hat{u}}_{k}} +  \frac{1}{2}\overline{\hat{{\rho}}\hat{v}^{\prime \prime}(\hat{u}_{k}^{\prime \prime}\hat{u}_{k}^{\prime \prime})}. \label{eq:unclosed_pv_PE}
\end{align}
With the internal energy fluctuation $\hat{e}^{\prime \prime}$  obtained from \eqref{eq:int_energy}, \eqref{eq:unclosed_pv_PE} can be further expanded as
\begin{align}
\hat{R} \overline{\hat{\rho}\hat{v}^{\prime \prime}\hat{T}^{\prime \prime}} + \overline{\hat{\rho}\hat{v}^{\prime \prime} \hat{E}^{\prime \prime}} &= 
 \hat{R}\overline{\hat{\rho}\hat{v}^{\prime \prime}\hat{T}^{\prime \prime}} +  \hat{R}\overline{\hat{\rho}\hat{v}^{\prime \prime}\left(1.5 \hat{T}^{\prime \prime}-5226.625 \hat{Y}^{\prime \prime}_P-745.375\right)} + \overline{\hat{{\rho}}\hat{v}^{\prime \prime}\hat{u}_{k}^{\prime \prime}\widetilde{\hat{u}}_{k}} +  \frac{1}{2}\overline{\hat{{\rho}}\hat{v}^{\prime \prime}(\hat{u}_{k}^{\prime \prime}\hat{u}_{k}^{\prime \prime})} \notag \\
  &= 
 2.5{\hat{R}}\overline{\hat{\rho}\hat{v}^{\prime \prime}\hat{T}^{\prime \prime}}  -5226.625{\hat{R}}\overline{\hat{\rho}\hat{v}^{\prime \prime} \hat{Y}^{\prime \prime}_P} + \overline{\hat{{\rho}}\hat{v}^{\prime \prime}\hat{u}_{k}^{\prime \prime}\widetilde{\hat{u}}_{k}} +  \frac{1}{2}\overline{\hat{{\rho}}\hat{v}^{\prime \prime}(\hat{u}_{k}^{\prime \prime}\hat{u}_{k}^{\prime \prime})} \notag \\
 &= 
 2.5\frac{\gamma -1}{\gamma}\hat{C}_{p}\overline{\hat{\rho}\hat{v}^{\prime \prime}\hat{T}^{\prime \prime}}  -5226.625\frac{\gamma -1}{\gamma}\hat{C}_{p}\overline{\hat{\rho}\hat{v}^{\prime \prime} {\hat{Y}_P}^{\prime \prime}} + \overline{\hat{{\rho}}\hat{v}^{\prime \prime}\hat{u}_{k}^{\prime \prime}\widetilde{\hat{u}}_{k}} +  \frac{1}{2}\overline{\hat{{\rho}}\hat{v}^{\prime \prime}(\hat{u}_{k}^{\prime \prime}\hat{u}_{k}^{\prime \prime})} \notag\\
  &= 
 C_{1}\hat{C}_{p}\overline{\hat{\rho}\hat{v}^{\prime \prime}\hat{T}^{\prime \prime}}  -\hat{C}_{h2}\frac{(\gamma -1)}{\gamma}\hat{C}_{p}\overline{\hat{\rho}\hat{v}^{\prime \prime} {\hat{Y}_P}^{\prime \prime}} + \overline{\hat{{\rho}}\hat{v}^{\prime \prime}\hat{u}_{k}^{\prime \prime}\widetilde{\hat{u}}_{k}} +  \frac{1}{2}\overline{\hat{{\rho}}\hat{v}^{\prime \prime}(\hat{u}_{k}^{\prime \prime}\hat{u}_{k}^{\prime \prime})}. \label{eq:pp_vpp}
\end{align}
The constants are  $C_{1} = 2.5(\gamma -1)/{\gamma} = 1$ for argon ($\gamma = 1.667$) and $\hat{C}_{h2} = 5226.625\,\mathrm{K}$ (units of $\hat h/\hat R$) {for consistency with the thermodynamic polynomials \eqref{eq:enthap_reac}}.  Substituting \eqref{eq:pp_vpp} into \eqref{eq:dim_energy}, we obtain a second version of the unclosed RANS total energy equation,
\begin{align}
\frac{\partial (\bar{\hat{\rho}} \widetilde{\hat{E}})}{\partial \hat{t}} 
&+ \frac{\partial}{\partial \hat{y}} \left( \widetilde{\hat{v}} \bar{\hat{\rho}} \widetilde{\hat{E}} \right) 
+ \frac{\partial (\bar{\hat{p}} \widetilde{\hat{v}})}{\partial \hat{y}} 
- \frac{\partial}{\partial \hat{y}}\left(\frac{\hat{C}_{p}\hat{\mu}}{\Pr}  \frac{\partial \widetilde{\hat{T}}}{\partial \hat{y}}\right) 
- {\underbrace{\frac{\partial}{\partial \hat{y}}\left(\frac{\hat{C}_{p}\hat{\mu}}{\Pr} \frac{\partial \overline{\hat{T}^{\prime \prime}}}{\partial \hat{y}}\right)}_{\mathrm{S}_{n}}} \nonumber\\
&+ \frac{\partial}{\partial \hat{y}} \Bigg(  {\underbrace{\hat{C}_{p}\overline{\hat{\rho}\hat{v}^{\prime \prime}\hat{T}^{\prime \prime}}}_\mathrm{S3}  -\underbrace{\hat{C}_{h2}\frac{(\gamma -1)}{\gamma}\hat{C}_{p}\overline{\hat{\rho}\hat{v}^{\prime \prime} {\hat{Y}_P}^{\prime \prime}}}_\mathrm{S5}}  + {\underbrace{\widetilde{\hat{u}_k} \overline{\hat{\rho} \hat{u}_k^{\prime \prime} \hat{v}^{\prime \prime}}}_\mathrm{S6}} + {\underbrace{\frac{1}{2}\overline{\hat{\rho} \hat{v}^{\prime \prime} \hat{u}_k^{\prime \prime} \hat{u}_k^{\prime \prime}}}_\mathrm{S4}} \Bigg) \nonumber\\
&- \frac{\partial}{\partial \hat{y}} \Bigg( \sum_{k=1}^{N} D\bar{\hat{\rho}}\widetilde{\hat{h}}_k\frac{\partial \widetilde{\hat{Y}}_{k}}{\partial \hat{y}} + {\underbrace{\sum_{k=1}^{N}\overline{D\hat{\rho} \hat{h}_k^{\prime \prime}\frac{\partial \hat{Y}_k^{\prime \prime}}{\partial \hat{y}}}}_{\mathrm{S}_{n}}}  \Bigg) 
- \frac{\partial}{\partial \hat{y}} \big( {\underbrace{\overline{\hat{\tau}_{ij} \hat{u}_{i}^{\prime \prime}}}_\mathrm{S4'}} + \overline{\hat{\tau}_{ij}} \widetilde{\hat{u}}_{i} \big)  = 0, \label{eq:Appn:energy_unclosed}
\end{align}
where terms requiring closure are labeled with underbraces. 
The closures for these unclosed terms are outlined in Gatski and Bonnet \cite{gatski2013compressibility}, Wilcox \cite{wilcox1998turbulence}, and Hirsch \cite{hirsch2007numerical}.  The terms S$_n$ are typically neglected. 

The unclosed terms S3, S5, and S6 represent the turbulent transport of enthalpy due to temperature and species fluctuations, as well as the correlation of the Reynolds stress with the mean flow.  These are approximated using the gradient diffusion and Boussinesq hypotheses:
\begin{align}
\hat{C}_{p}\overline{\hat{\rho}\hat{v}^{\prime \prime}\hat{T}^{\prime \prime}}   &\approx -\frac{\hat C_{p}\hat\mu_{t,E}(Q;\theta)}{\Pr_{t}}\frac{\partial \widetilde{\hat{T}}}{\partial \hat{y}}\label{eq:unclosed_term_S3},\\
\hat{C}_{h2}\frac{\gamma -1}{\gamma}\hat{C}_{p}\overline{\hat{\rho}\hat{v}^{\prime \prime} {\hat{Y}_P}^{\prime \prime}}  &\approx -\hat{C}_{h2}\frac{(\gamma - 1)}{\gamma}{\hat C_{p}}\frac{\hat\mu_t(Q;\theta)}{\Sc_{t}}\frac{\partial \widetilde{\hat{Y}}_{P}}{\partial \hat{y}},\label{eq:unclosed_term_S5}\\
{\widetilde{\hat{u}}_k \overline{\hat{\rho} \hat{u}_k^{\prime \prime} \hat{v}^{\prime \prime}}} &\approx -\left(\widetilde{\hat{u}}\hat\mu_t(Q;\theta)\left(\frac{\partial \hat{u}}{\partial \hat{y}}\right) + \frac{4}{3}\widetilde{\hat{v}}\hat\mu_t(Q;\theta)\left(\frac{\partial \hat{v}}{\partial \hat{y}}\right)\right) + \frac{2}{3}\bar{\hat{\rho}}\hat{k}\label{eq:unclosed_term_S6},
\end{align}
where we introduce the turbulent thermal conductivity $\hat C_p\hat\mu_{t,E}/\Pr_t$. 

The terms S4 and S4$'$ are the turbulent kinetic energy flux and the viscous work performed by the fluctuations; these are conventionally closed together using a single gradient-diffusion model:
\begin{equation}
\label{eq:unclosed_term_S7}
    \frac{1}{2}\overline{\hat{\rho} \hat{v}^{\prime \prime} \hat{u}_k^{\prime \prime} \hat{u}_k^{\prime \prime}} - \overline{\hat{\tau}_{ij} \hat{u}_{i}^{\prime \prime}}  \approx -\left(\hat{\mu} + \frac{\hat\mu_t(Q;\theta)}{\sigma_k}\right) \frac{\partial \hat{k}}{\partial \hat{y}}.
\end{equation}
Substituting  the closures from \eqref{eq:unclosed_term_S5}, \eqref{eq:unclosed_term_S6},  and \eqref{eq:unclosed_term_S7} into \eqref{eq:Appn:energy_unclosed} and nondimensionalizing yields the final form of the RANS total energy equation:
\begin{align}
\label{eq:appen_final_deriv}
\frac{\partial (\bar{\rho} \widetilde{E})}{\partial t}  
  &+ \frac{\partial}{\partial y} \left( \widetilde{v} \bar{\rho} \widetilde{E} \right) + \frac{1}{\Ma^2}\frac{\partial (\bar{p} \widetilde{v})}{\partial y} - \frac{1}{\Ma^2 \Rey {\Pr} } \frac{\partial}{\partial {y}}\left({{{\mu}}} \frac{\partial \widetilde{{T}}}{\partial {y}}\right) - \underbrace{{\frac{1}{\Ma^2 \Rey {\Pr_{t}}} \frac{\partial}{\partial {y}}\left({{\mu_{t,E}(Q;\theta)}} \frac{\partial \widetilde{{T}}}{\partial {y}}\right)}}_\mathrm{S3}  \nonumber \\
  &- \frac{1}{\Rey } \frac{\partial}{\partial {y}} \left( \widetilde{{u}}\mu\frac{\partial {\widetilde{u}}}{\partial {y}} + \frac{4}{3}\widetilde{{v}}\mu\frac{\partial {\widetilde{v}}}{\partial {y}} \right) - \underbrace{\frac{1}{\Rey}\frac{\partial}{\partial y} \left( \left(\mu + \frac{\mu_t(Q;\theta)}{\sigma_k}\right) \frac{\partial k}{\partial y} \right)}_\mathrm{S4}
  - \frac{1}{\Ma^2 \Rey \Sc \gamma} \frac{\partial}{\partial {y}} \left( \rho \mu \sum_{k=1}^{N} \widetilde{h}_k\pp{\widetilde{Y}_{k}}{y} \right)  \nonumber \\
  &+ \underbrace{\frac{C_{h3}}{\Ma^2\Rey\Sc_{t}\gamma}{\frac{\partial}{\partial y}\left({\mu_t(Q;\theta)}\frac{\partial \widetilde{Y}_{P}}{\partial {y}}\right)}}_\mathrm{S5} 
  - \underbrace{ \frac{1}{\Rey }\frac{\partial}{\partial {y}} \left( \widetilde{{u}}\mu_t(Q;\theta)\frac{\partial {\widetilde{u}}}{\partial {y}} + \frac{4}{3}\widetilde{{v}}\mu_t(Q;\theta)\frac{\partial {\widetilde{v}}}{\partial {y}} \right) + \pp{}{y}\left(\frac{2}{3}\widetilde{{v}}\bar{{\rho}}{k}\right)}_\mathrm{S6} = 0.
\end{align}
In this final step,  $C_{h3} = 5226.625/\hat{T}_{0} = 17.422$ is the dimensionless version of $\hat{C}_{h2}$, where $\hat{T}_{0} = 300$ K. This results in the closed form of the total energy transport equation, \eqref{eq:rans_E} in the text.

\section{Training Convergence}
\label{sec_appen: training_convergence}

\begin{figure}
    \centering
    \includegraphics[width=0.49\linewidth]{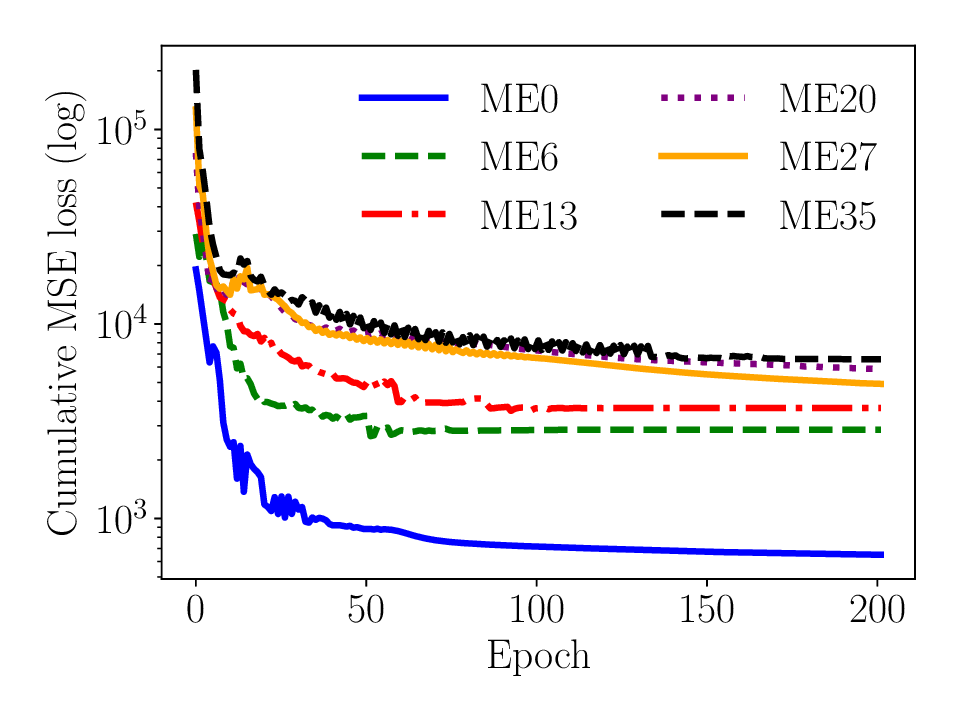} 
    \hfill 
    \includegraphics[width=0.49\linewidth]{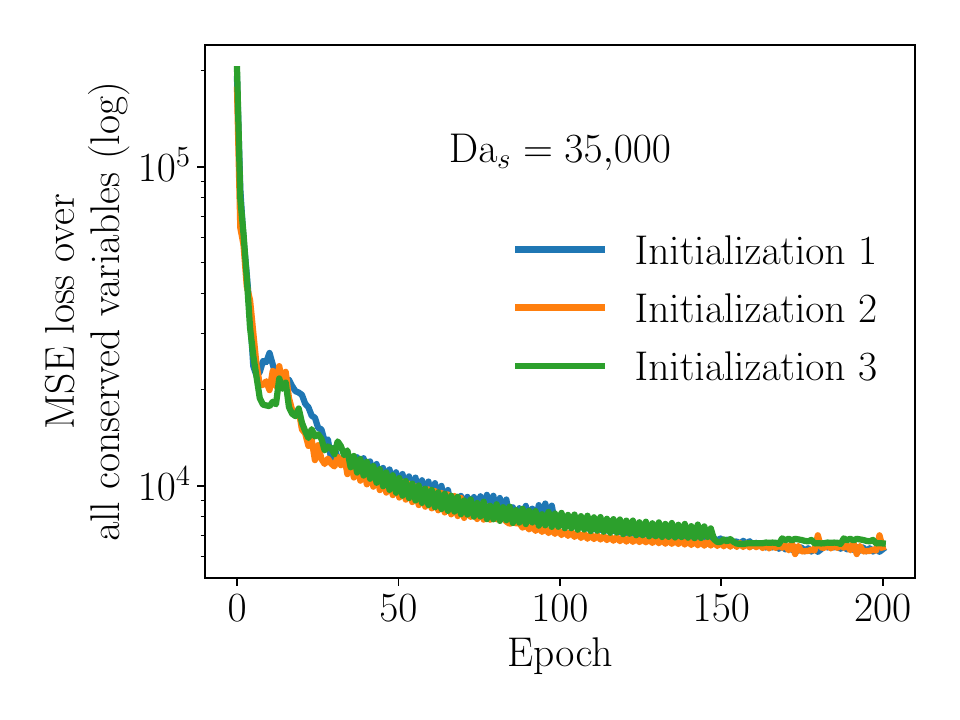}  
     \caption{Convergence of the training loss versus epochs for single-\(\Da_{s}\) training (left) and training loss convergence over time for \(\Da_{s} = 35{,}000\) with three distinct parameter initialization (right).}
    \label{fig:train_loss_convergence}
\end{figure}

Figure~\ref{fig:train_loss_convergence} presents the objective-function training convergence for different single-\(\Da_{s}\) models using identical random neural network initializations. Notably, all models achieve over 90\% loss reduction within 200 epochs, with the $\Das=0$ model achieving over two orders of magnitude convergence. This confirms the training convergence of the trained models.

We further assess the sensitivity of our approach to parameter initialization. Figure \ref{fig:train_loss_convergence} shows the objective-function convergence for three neural networks initialized using different random seeds and trained for $\Da_{s} = 35{,}000$. The convergence of the training loss is similar across all three initializations, thus the uncertainty of the MSE loss would be relatively low. However, fully quantifying this uncertainty would be computationally expensive, as it would necessitate significantly more training initializations. At a minimum, this similarity confirms the stability of the training process.


\end{document}